\documentclass[12pt]{article}

\usepackage[dvips]{graphicx}
\usepackage{epsfig}
\usepackage{amsmath,amsfonts,amssymb,amsthm}
\usepackage{verbatim}
\usepackage{psfrag}
\usepackage{bm}
\usepackage{bbm}
\usepackage[square,comma,sort&compress,numbers]{natbib}
\usepackage{color}
\usepackage{slashed}
\usepackage{upgreek}

\usepackage{epsf,epsfig}
\usepackage{graphics}

\newcommand{\gsim}
{\;\raisebox{-.3em}{$\stackrel{\displaystyle >}{\sim}$}\;}

\newcommand{\beq}{\begin{equation}}
\newcommand{\eeq}{\end{equation}}
\newcommand{\bea}{\begin{eqnarray}}
\newcommand{\eea}{\end{eqnarray}}

\newcommand{\bce}{\begin{center}}
\newcommand{\ece}{\end{center}}

\begin{document}
\thispagestyle{empty}

\begin{flushright}
{
\small
TUM-HEP-947-14\\
NSF-KITP-14-061
}
\end{flushright}

\vspace{0.4cm}
\begin{center}
\Large\bf\boldmath
Strong Washout Approximation to Resonant Leptogenesis
\unboldmath
\end{center}

\vspace{0.4cm}

\begin{center}
{Bj\"orn~Garbrecht, Florian Gautier and Juraj Klaric}\\
\vskip0.2cm
{\it Physik Department T70, James-Franck-Stra{\ss}e,\\
Technische Universit\"at M\"unchen, 85748 Garching, Germany}\\
\vskip1.4cm
\end{center}

\begin{abstract}
We show that the effective decay asymmetry for resonant Leptogenesis in 
the strong washout regime with two
sterile neutrinos and a single active flavour can in wide regions of parameter
space be approximated by its late-time limit $\varepsilon=X\sin(2\varphi)/(X^2+\sin^2\varphi)$,
where $X=8\pi\Delta/(|Y_1|^2+|Y_2|^2)$, $\Delta=4(M_1-M_2)/(M_1+M_2)$,
$\varphi=\arg(Y_2/Y_1)$,
and $M_{1,2}$, $Y_{1,2}$ are the masses and Yukawa couplings of the sterile
neutrinos.
This approximation in particular extends to parametric regions where $|Y_{1,2}|^2\gg \Delta$,
{\it i.e.} where the width dominates the mass splitting.
We generalise the formula for the effective decay asymmetry to the case of several
flavours of active leptons and demonstrate how this quantity can be used to
calculate the lepton asymmetry for phenomenological scenarios that are in agreement with
the observed neutrino oscillations.
We establish analytic criteria for the validity of the late-time approximation
for the decay asymmetry and
compare these with numerical results that are obtained by solving for the mixing and
the oscillations of the sterile neutrinos. For 
phenomenologically viable models with two sterile neutrinos, we find that
the flavoured effective late-time decay asymmetry can be applied throughout parameter space.
\end{abstract}



\section{Introduction}

Resonant enhancement from mass degeneracies is a way
of obtaining sizeable charge-parity ($CP$) violating effects, that
would be strongly suppressed by powers of small couplings otherwise.
Depending on the ratio of the mass splitting to the decay rate in
a system of mixing particles, it may either be more advantageous to describe
the $CP$-violating effects as a time-dependent phenomenon due to mixing and
oscillations of the almost mass-degenerate states, or, further away from the
mass degeneracy, in terms of a time-independent effective decay
asymmetry~\cite{Beringer:1900zz}. The important role that resonant $CP$-violation
assumes in many systems that can be tested in the laboratory has lead to
the idea that a resonantly enhanced decay asymmetry for sterile neutrinos
may have been of importance for Leptogenesis in the Early Universe~\cite{Covi:1996wh,Flanz:1996fb,Pilaftsis:1997dr,Pilaftsis:1997jf,Pilaftsis:2003gt,Pilaftsis:2005rv}.

Standard Leptogenesis calculations typically rely on a time-independent effective asymmetry
$\varepsilon$, which may be resonantly enhanced or not.
It isolates the $CP$-violating loop effects from
the leading-order out-of-equilibrium dynamics, that may be described in terms
of tree-level
rates, see {\it e.g.} Refs.~\cite{Giudice:2003jh,Buchmuller:2004nz}.
While this separation approach brings along some caveats and pitfalls, most notably
the necessity of a subtraction of real intermediate states (RIS) in order to comply with
the consequences of the
combined charge-, parity- and time-reversal symmetry~\cite{Kolb:1979qa}, it has proved
very useful for practical phenomenological calculations as well as
for the conceptual description of
the dynamics of the generation and the freeze out of
the lepton asymmetry.

A more unified approach to Leptogenesis, starting from first principles, is provided
by the Closed-Time-Path (CTP) method~\cite{Schwinger:1960qe,Keldysh:1964ud,Calzetta:1986cq},
that is formulated in terms of Green functions
and leads to kinetic equations that readily encompass the crucial higher-order
corrections~\cite{Buchmuller:2000nd,De Simone:2007rw,Garny:2009rv,Garny:2009qn,Anisimov:2010aq,Garny:2010nj,Beneke:2010wd,Beneke:2010dz,Garny:2010nz,Anisimov:2010dk}.
No ad hoc subtraction of RIS is needed here. For the present context,
we note that in the appropriate limiting cases, we recover either
the description of resonant Leptogenesis from mixing and oscillations or
in terms of a time-independent decay asymmetry $\varepsilon$~\cite{Garbrecht:2011aw}.
Both regimes overlap, such that suitable calculational
methods for all parametric configurations
are available. A general formulation that spans from the regime where the
sterile neutrinos are fully relativistic to the case when these are
non-relativistic, which is of relevance for strong washout
and that accounts for the expansion of the Universe, is developed in
Refs.~\cite{Garbrecht:2011aw,Drewes:2012ma},
that are a main basis for the present work. The derivation in Refs.~\cite{Garbrecht:2011aw,Drewes:2012ma} relies on
Green functions in Wigner-space (where the two-point functions are Fourier
transformed with respect to the relative coordinate). Alternative approaches
also based on the CTP method employ Green functions in the coordinate
representation~\cite{Garny:2011hg,Iso:2013lba,Iso:2014afa,Hohenegger:2014cpa},
but the results agree with those obtained in Wigner space, which is most evident
when comparing Refs.~\cite{Garbrecht:2011aw} and~\cite{Iso:2014afa}, where consistent
effective evolution
equations for the sterile neutrinos and for the final freeze-out
asymmetry are obtained.

We also note that mixing and oscillations can be treated within a density matrix approach,
that is typically applied to Leptogenesis in the fully relativistic regime,
see Refs.~\cite{Akhmedov:1998qx,Asaka:2005pn,Shaposhnikov:2008pf,Canetti:2010aw,Canetti:2012vf,Canetti:2012zc,Canetti:2012kh,Shuve:2014zua}. More recently, the density matrix method has also been applied
to Leptogenesis in the non-relativistic strong washout
regime~\cite{Dev:2014laa}.

While the CTP formulation of resonant Leptogenesis is rederived 
and confirmed in Ref.~\cite{Iso:2014afa},
an important point concerning approximate solutions is added there: Since by
definition of the strong washout regime, the relaxation rate $\Gamma$
of the sterile neutrinos exceeds the Hubble rate $H$, neglecting time-derivatives acting
on the non-equilibrium distributions of the sterile neutrinos should only incur an
error that is of order $H/\Gamma$. This allows for a quasi-static solution
for the right-handed neutrino distributions and their off-diagonal correlations,
from which an effective late-time decay-parameter $\varepsilon$ can be constructed,
even when their mass splitting is smaller than their decay rate.

Based on above developments, we present here the following points that are of
relevance for resonant Leptogenesis in the strong washout regime:
\begin{itemize}
\item
We show how the non-relativistic approximations and simplifications,
that are of relevance in the strong washout
regime, follow from the general treatment of
Refs.~\cite{Garbrecht:2011aw,Drewes:2012ma}.
\item
We define the effective decay asymmetry $\varepsilon$ as the lepton asymmetry
that results on average from the decay of one out-of-equilibrium
sterile neutrino. When compared to the decay asymmetry introduced in
Ref.~\cite{Iso:2014afa}, this definition resembles more closely the expressions
that are usually employed in Leptogenesis calculations, such that it leads to a simple and
straightforward way of obtaining the lepton asymmetry.
We present the relevant equations that determine the freeze-out asymmetry as
well as example solutions.
\item
We give an expression for the decay asymmetry taking account of
active lepton flavours and their possible correlations. We emphasise
that flavour effects
should be phenomenologically relevant throughout the
parameter space. Again, we illustrate the use of this effective asymmetry with
numerical examples.
\item
Since it is crucial for resonant Leptogenesis to treat the decay rate $\Gamma$
of the sterile neutrinos
as matrix-valued, the criterion $H/\Gamma\ll 1$ for the applicability of the
approximation in terms of an effective decay asymmetry can only be of schematic
meaning. For a simplified scenario with one active lepton flavour only,
we determine the smallest eigenvalue associated with the linear differential
equation that governs the evolution of the
sterile neutrino densities and their flavour-off-diagonal correlations.
By comparison with the Hubble rate,
this eigenvalue can be used in order to assess whether the approximation
in terms of the effective decay asymmetry $\varepsilon$ is applicable.
\item
For a phenomenological scenario with two sterile neutrinos,
that explains the observed oscillations of active neutrinos,
we find that the use of
the effective late-time decay-asymmetry can be justified for
all regions of parameter space. This conclusion is also based on
comparing the eigenvalues of the equations that govern the mixing
and the oscillations of the sterile neutrinos with the Hubble expansion rate
prior to the freeze out of the lepton asymmetry.
\end{itemize}

\section{Relativistic Resonant Leptogenesis}

We consider the usual see-saw model for neutrino masses that is given by
the Lagrangian
\begin{align}
\label{Lagrangian}
{\cal L}=\frac{1}{2}\bar{N}_i({\rm i} \partial\!\!\!/-M)_{ij} N_j
+\bar{\ell}_a{\rm i}\partial\!\!\!/\ell_a
+(\partial^\mu\phi^\dagger)(\partial_\mu \phi)
-Y_{ia}^*\bar\ell_a \epsilon_{{\rm SU}(2)} \phi P_{\rm R}N_i
-Y_{ia}\bar N_iP_{\rm L}\phi^\dagger\epsilon_{{\rm SU}(2)}^\dagger\ell_a\,.
\end{align}
Here, the $N_i$ are the sterile neutrinos, that observe the Majorana condition
$N_i^c=N_i$, where the superscript $c$ stands for charge conjugation. The Higgs doublet is given
by $\phi$ and $\epsilon_{{\rm SU}(2)}$ is the antisymmetric, ${\rm SU}(2)$-invariant
tensor with $\epsilon_{{\rm SU}(2)}^{12}=1$. The Standard Model (SM) lepton doublets are given
by $\ell_a$, where $a=e,\mu,\tau$. When considering the single-flavour model, we
drop the index a on the fields $\ell$ as well as the on Yukawa couplings $Y$. We make
use of the freedom of field redefinitions in order to choose the symmetric
matrix $M$ to be real and diagonal, and we refer to the diagonal elements as
$M_i\equiv M_{ii}$.

We describe the generation of the comoving lepton charge density $q_{\ell ab}$ in terms
of a source term $S_{ab}$ and a washout term $W$ as~\cite{Garbrecht:2011aw,Drewes:2012ma}
\begin{align}
\label{Beq:lepton}
q_{\ell ab}^\prime=g_w S_{ab}-\frac12\{W,q_\ell\}_{ab}\,.
\end{align}
The charge density accounts for the gauge multiplicity, hence we include
here the factor $g_w=2$. Moreover, as mentioned in the Introduction, we
allow for the possibility of correlations of the SM lepton flavours. 
The expansion of the Universe is accounted
for through the metric in conformal coordinates
$g_{\mu\nu}=a(\eta)\eta_{\mu\nu}$, where $\eta_{\mu\nu}$ is the Minkowski
metric, $a(\eta)$ is the scale factor
and $\eta$ is conformal time. A prime denotes a derivative
with respect to $\eta$.

In Ref.~\cite{Garbrecht:2011aw}, it is shown that the source term
for resonant Leptogenesis through the lepton-number violating Majorana
mass
can be computed by first solving for the flavour correlations of the
oscillating sterile neutrinos, similar to the standard calculations for $CP$-violation
in mixing meson systems~\cite{Beringer:1900zz} or to the lepton-number conserving source
in the scenarios that are usually referred to as Leptogenesis from
neutrino oscillations~\cite{Akhmedov:1998qx,Asaka:2005pn,Shaposhnikov:2008pf,Canetti:2010aw,Canetti:2012vf,Canetti:2012zc,Canetti:2012kh,Shuve:2014zua,Drewes:2012ma}.
The result of Ref.~\cite{Garbrecht:2011aw} is
generalised to include flavour correlations in Ref.~\cite{Drewes:2012ma}
and then reads
\begin{align}
\label{sourceterm:flavoured}
S_{ab}=-\sum\limits_{\overset{i,j}{i\not=j}} Y_{ia}^* Y_{jb} \int\frac{d^4 k}{(2\pi)^4}
{\rm tr}
\left[
P_{\rm R}{\rm i}\delta S_{N ij}(k) 2P_{\rm L}
\hat{\slashed\Sigma}^{\cal A}_N(k)
\right]\,,
\end{align}
where $\hat{\slashed\Sigma}^{\cal A}_N(k)$ is the reduced spectral self-energy
of the sterile neutrinos as defined in Ref.~\cite{Drewes:2012ma}.
The correlations of the sterile neutrinos are described by
${\rm i}\delta S_{N ij}(k)$.
Besides the indices $i,j$ for the sterile neutrino flavours, this function corresponds
to a rank two tensor in terms of Dirac spinors. It satisfies
Kadanoff-Baym equations and the solutions can be decomposed as
\begin{align}
\label{helicity:decomposition}
{\rm i}\delta S_N=\sum\limits_{h=\pm}{\rm i}\delta S_{Nh}\,,
\qquad
-{\rm i}\gamma^0 \delta S_{Nh}=\frac14(\mathbbm 1+h \hat k^i \sigma^i)
\otimes \rho^a g_{ah}\,,
\end{align}
where $\sigma$ and $\rho$ are Pauli matrices.
In the resonant regime $|M_i-M_j| \ll \bar M $,
the different components may be written as~\cite{Garbrecht:2011aw}
\begin{align}
g_{ahij}(k)=2\pi\delta(k^2-a^2 \bar M^2) 2k^0 \delta f_{ahij}\,,
\end{align}
where $\bar M=(M_i+M_j)/2$. Moreover, the
Kadanoff-Baym equations also imply the relations~\cite{Garbrecht:2011aw} 
\begin{align}
\label{constrf1f3f0}
\delta f_{1hij}(k)=\delta f_{3hij}(k)a\frac{M_{i}+M_{j}}{2h|\mathbf k|}\,,\qquad
\delta f_{1hij}(k)=\delta f_{0hij}(k)a\frac{M_{i}+M_{j}}{2k^0}\,.
\end{align}
In view of the non-relativistic approximation below, the $a=0$ component
is of particular interest.
The function $\delta f_{0hij}$ may be interpreted as the distribution
function of the sterile neutrinos and of their flavour correlations.
Using the decomposition~(\ref{helicity:decomposition}) and the relations~(\ref{constrf1f3f0}),
the source term~(\ref{sourceterm:flavoured}) can be expressed as
$S_{ab} \equiv  \int\frac{d^3 k}{(2\pi)^3} \mathcal{S}_{ab}(\mathbf{k})$,
where
\begin{align}
\label{sourceterm:flavoured_bis}
\mathcal{S}_{ab}(\mathbf{k}) = \sum_{\overset{i,j}{i\not=j}}\sum_{h = \pm} Y_{ia}^{*} Y_{jb}\bigg\{& \frac{k\cdot\hat\Sigma^{\cal A}_{N}(k)}{k^0} \left[ \delta f_{0hij}(k)-  \delta f^{*}_{0hij}(k) \right] 
\\\notag
+& h \frac{\tilde k\cdot\hat\Sigma^{\cal A}_{N}(k)}{k^0} \left[ \delta f_{0hij}(k)+  \delta f^{*}_{0hij}(k)\right] \bigg\} \bigg|_{k^0 = \omega(\mathbf{k})}\,,
\end{align}
$\omega(\mathbf{k})= \sqrt{\mathbf{k}^2+ a \bar M^2}$,
$\tilde k={(|\mathbf k|,k^0 \mathbf{k}/|\mathbf{k}|)}$
and
$\delta f^{*}_{0hij}(k^0) = \delta f_{0hij}(-k^0) $. 
The Kadanoff-Baym 
equations imply that the sterile neutrino
distributions and their correlations
satisfy~\cite{Garbrecht:2011aw,Iso:2014afa}
\begin{align}
\label{sterile:oscillations}
\delta  f^\prime_{0h}+\frac{a^2(\eta)}{2k^0}{\rm i}[M^2,\delta f_{0h}]+f^{{\rm eq}^\prime}
=&
- g_w \bigg\{{\rm Re}[Y^* Y^t]\frac{k\cdot\hat\Sigma^{\cal A}_{N}}{k^0} - {\rm i}h{\rm Im}[Y^* Y^t]\frac{\tilde k\cdot\hat\Sigma^{\cal A}_{N}}{k^0},\delta f_{0h}\bigg\}
\,,
\end{align}
where $f^{\rm eq}$ is the equilibrium Fermi-Dirac distribution of the sterile neutrinos.
One may alternatively derive this equation using a more heuristic approach in terms
of a density matrix instead of the two-point function of the sterile neutrinos.
The solution may be substituted back into the source term~(\ref{sourceterm:flavoured})
and eventually into the equation for generating the lepton
charge-density~(\ref{Beq:lepton})
in order to obtain predictions for the freeze-out asymmetry.

When comparing Eq.~(\ref{sterile:oscillations}) with the correpsonding
expressions in {\it e.g.} Ref.~\cite{Cirigliano:2009yt} (for oscillations
of scalar particles derived in the CTP framework) or~\cite{Sigl:1992fn} (for
neutrino oscillations using a density-matrix approach) one notices that
the commutator term in these references involves a matrix of
frequencies $\omega$ rather than $M^2$. The different forms are
consistent in the resonant regime where
$M^2_i-M^2_j\ll \bar\omega^2=\mathbf k^2 + \bar M^2$ because there
is agreement to leading order in $(M^2_i-M^2_j)/\bar \omega^2$:
$\omega=\sqrt{\mathbf k^2 + M^2}=\bar\omega + \delta M^2/(2\bar\omega)+{\cal O}\left([(M^2_i-M^2_j)/\bar \omega^2]^2\right)$, where
we have
written $M^2=\bar M^2+\delta M^2$. The commutator, of course, only
depends on the non-diagonal terms, such that
$[2\bar\omega^2+\delta M^2,\cdot]\equiv[\delta M^2,\cdot]\equiv[M^2,\cdot]$.
While the derivation in Ref.~\cite{Cirigliano:2009yt} relies on approximations
up to ${\cal O}\left((M^2_i-M^2_j)/\bar \omega^2\right)$, it is demonstrated in
Ref.~\cite{Fidler:2011yq} using the CTP approach that the form with $\omega$ in
the commutator indeed corresponds to the correct kinetic term to all orders. However,
one should be aware of the fact that the collison
term on the right hand side of Eq.~(\ref{sterile:oscillations}) is evaluated
to order $[(M^2_i-M^2_j)/\bar \omega^2]^0$ only. Extending to higher orders requires
a gradient expansion of the convolution of Wigner functions, which is formally
worked out also in Ref.~\cite{Fidler:2011yq}, but leads to considerable complications.
In conlcusion, the present form of the commutator term is not only a sufficiently accurate
approximation for the present purposes, but a consistent treatment to higher orders
in $[(M^2_i-M^2_j)/\bar \omega^2]$ would also imply a considerably more complicated form
of the collision term. This has neither been worked out yet in the context of resonant
Leptogenesis, nor is this necessary in order to obtain results to leading accuracy.

\section{Non-Relativistic Approximations}

Now, we consider a situation, where $\bar M\gg T$ (and all sterile neutrinos
are assumed to be close together in mass, $|M_i-M_j|\ll\bar M$),
as it is of relevance in strong washout
scenarios around the time of freeze out.
The main simplification arises here due to the fact that modes that do not
satisfy $|\mathbf k|\ll aM$ are strongly Maxwell suppressed, such that we
may approximate the four momenta as
\begin{align}
k^\mu=(k^0,\mathbf k)\approx(\pm a \bar M,\mathbf 0)\,,\quad
\tilde k^\mu\approx(0,k^0 \mathbf k/|\mathbf k|)\,.
\end{align}
Due to the same reason, we can neglect the thermal contributions to the spectral self-energy
of the sterile neutrinos, such that it takes its vacuum form
\begin{align}
\left(\hat\Sigma_N^{\cal A}\right)^\mu={\rm sign}(k^0) \frac{k^\mu}{32\pi}\,.
\end{align}

For the terms involving $\hat\Sigma_N^{\cal A}$
that appear in Eq.~(\ref{sterile:oscillations}), this implies that we can take the
approximate forms
\begin{align}
k\cdot\hat\Sigma_N^{\cal A}={\rm sign}(k^0)\frac{a^2 \bar M^2}{32\pi}\,,\qquad
\tilde k\cdot \hat\Sigma^{\cal A}_N=0\,.
\end{align}
Then, we integrate that equation with the result
\begin{align}
\label{sterile:density:oscillations}
\delta  n^{\pm\prime}_{0h}\pm\frac{a}{2\bar M}{\rm i}[M^2,\delta n^\pm_{0h}]+n^{{\rm eq}^\prime}
=&
- \frac{g_w a \bar M}{32\pi} \bigg\{{\rm Re}[Y^* Y^t],\delta n^\pm_{0h}\bigg\}
\,,
\end{align}
where we have defined
\begin{align}
\delta n^\pm_{0h}=\int\frac{d^3k}{(2\pi)^3}\delta f_{0h}(\pm \omega(\mathbf k),\mathbf k)\,.
\end{align}
This is the comoving non-equilibrium
number density of sterile neutrinos, $\delta n^\pm_{0hij}=\delta n^{\pm *}_{0hji}$, which
is of the form of a  Hermitian matrix.
The comoving equilibrium number density is denoted by $n^{\rm eq}$.
The Majorana nature of the sterile neutrinos implies that
$\delta n^+_{0h ij}=\delta n^{-*}_{0h ij}$, a property that is directly inherited from
the distribution $\delta f_{0h}(\pm \omega,\mathbf k)$ and that is derived
in Ref.~\cite{Garbrecht:2011aw}. Note that in the non-relativistic limit,
the solutions for the sterile neutrino densities are helicity independent. The relativistic
generalisation that accounts for helicity is worked out in Ref.~\cite{Garbrecht:2011aw}.

In order to substitute these results into the source term~(\ref{sourceterm:flavoured}),
we use the relations~(\ref{constrf1f3f0})
that imply a vanishing axial density $\delta f_{3hij}$ in the non-relativistic limit. Note
moreover that the Dirac trace in Eq.~(\ref{sourceterm:flavoured}) selects then contributions
from $\delta f_{0h}$ only. The result for the flavoured source term in the non-relativistic
approximation then is
\begin{align}
\label{source:fl:NR}
S_{ab}=\frac{a\bar M}{16\pi}\sum\limits_{\overset{i,j}{i\not=j}}
Y^*_{ia}Y_{jb} \left(\delta n^+_{0hij}-\delta n^-_{0h ij}\right)\,.
\end{align}
Note that we do not sum over $h$ here and make use of the fact that in the non-relativistic
limit, we can approximate $n^\pm_{0+ij}=n^\pm_{0-ij}$.

\section{Strong Washout Regime}

In the radiation-dominated Universe, $a(\eta)=a_{\rm R}\eta$.
A particularly convenient choice is $\eta=1/T$, what requires
$a_{\rm R}=m_{\rm Pl}\sqrt{45/(4g_\star \pi^3)}\equiv T^2/H$. Moreover, one can then
easily define the parameter $z=\bar M/T=\bar M \eta$, that is often used in Leptogenesis
calculations.

We investigate under which circumstances the maximal enhancement of the decay
asymmetry can be attained.
For this purpose, we solve the Eq.~(\ref{sterile:density:oscillations}) in the
form that is obtained when using
above parametrisation in terms of $z$
\begin{align}
\label{nosc:cosmic}
\bar M\frac{d}{dz}\delta n_{0h}^\pm
\pm\frac{{\rm i}a_{\rm R}z}{2 \bar M^2}[M^2,\delta n_{0h}^\pm]+a_{\rm R}z\frac12\bar\Gamma\{{\rm Re}[ Y^* Y^t],\delta n_{0h}^\pm\}
+\bar M\frac{d}{dz}n^{\rm eq}=0\,,
\end{align}
where
\begin{align}
n^{\rm eq}=2^{-\frac32}\pi^{-\frac32}z^{\frac32}{\rm e}^{-z}a_{\rm R}^3\times{\rm diag}(1,1)
\end{align}
and $\bar\Gamma=1/(8\pi)$.
Since larger entries of $Y$ correspond to larger washout, it is proposed in
Ref.~\cite{Iso:2014afa} to obtain a simplified approximation in the strong washout
regime by neglecting the first term of Eq.~(\ref{nosc:cosmic}).
To put this more precisely, note that
out of the first three terms of Eq.~(\ref{nosc:cosmic}), which
are the homogeneous terms,
the second and the third grow with $z$. Therefore, neglecting the first term corresponds
to taking the late-time limit of the solution. If the late time-limit applies before
the freeze-out of the lepton asymmetry, that occurs for $z=z_{\rm f}$, it
leads to a valid approximation of the freeze-out asymmetry.

It is conceptually interesting to include also thermal masses for
the sterile neutrinos in addition
to the Majorana masses within Eq.~(\ref{nosc:cosmic}). In the non-relativistic
regime, the thermal
mass squares are of order $YY^\dagger T^2$, which is to be compared
with $M$ times the width of the sterile neutrinos, what is of order
$YY^\dagger M^2$. We therefore neglect this effect in the present
context where we can assume that $M\gg T$ and refer to
Ref.~\cite{Hohenegger:2014cpa}, where details on how to include thermal masses of
the sterile neutrinos are worked out.

The evolution of the lepton asymmetry is governed by the equation
\begin{align}
\label{BEq:ell}
-\bar M \frac{d}{dz}  \Delta_{\ell ab}=&g_w S_{ab}-\frac12 \{W,q_{\ell}\}_{ab}
-\frac12 W_{ab} q_\phi-\Gamma^{\rm fl}_{\ell ab}\\\notag
\equiv&4\varepsilon_{ab}(z)\bar M\frac{d}{dz} n^{\rm eq}-\frac12 \{W,q_{\ell}\}_{ab}
-\frac12 W_{ab} q_\phi-\Gamma^{\rm fl}_{\ell ab}
\,,
\end{align}
where the last equality defines the time-dependent
effective decay asymmetry $\varepsilon_{ab}(z)$,
in consistency with Eq.~(\ref{epsilon:effective}) below.
In view of flavour effects, we have written this in terms of the asymmetries
$\Delta_{\ell aa}=B/3-q_{\ell aa}$
that are conserved by SM interactions and where $B$ is
the baryon number density. Off-diagonal flavour-correlations
can be accounted for by $\Delta_{\ell ab}=-q_{\ell ab}$ for $a\not=b$, if necessary.
Moreover, $q_\phi$ stands for the
charge density in Higgs bosons, that is present in general.
We have also expressed Eq.~(\ref{BEq:ell}) in a way that
defines the decay asymmetry $\varepsilon$ as the the lepton asymmetry that results from
one sterile neutrino that initially drops out of equilibrium as a mass eigenstate. Note that
the factor of four in front of $\varepsilon_{ab}$ arises because of the two helicity
eigenstates of to the two sterile neutrinos. In addition,
this equation includes the crucial washout term $W$ in its flavoured variant, that is
derived in Ref.~\cite{Beneke:2010dz}\footnote{Here, we define it in a different
manner such that it is larger by a factor of two compared to its form
in Ref.~\cite{Beneke:2010dz}.}, see also Refs.~\cite{Blanchet:2011xq,Dev:2014laa}.
In the present context, we are interested in the situation where the sterile neutrinos are non-relativistic,
such that the washout matrix can be approximated by
\begin{align}
W=Y^\dagger Y \frac{3 a_{\rm R}}{2^\frac{7}{2}\pi^\frac{5}{2}}z^\frac{5}{2}{\rm e}^{-z}\,.
\end{align}
Lepton-flavour violating interactions mediated through SM Yukawa-couplings
are described by the term $\Gamma^{\rm fl}_{\ell ab}$,
that is defined and explained in Ref.~\cite{Beneke:2010dz}. In the fully flavoured approximation,
one assumes that these interaction delete the off-diagonal correlations in $q_\ell$ and $\Delta$.
Effectively, one may then just set the off-diagonal elements to zero and ignore
$\Gamma^{\rm fl}_{\ell ab}$.

Solving Eq.~(\ref{nosc:cosmic}) when neglecting
the derivatives acting on $\delta n^\pm_{0h}$ yields
for the off-diagonal correlations ($i\not=j$) of the sterile neutrinos
\begin{align}
\label{deltan:latetime}
\delta n_{0h ij}=&\frac{\bar M}{2 D}([YY^\dagger]_{ij}+[Y^*Y^t]_{ij})
([YY^\dagger]_{ii} +[YY^\dagger]_{jj})
\\\notag
\times&
[\bar M^2 \bar\Gamma([YY^\dagger]_{ii} +[YY^\dagger]_{jj})-{\rm i}(M_i^2-M_j^2)]
\times\frac{\bar M^2}{a_{\rm R} z}\frac{d}{dz} n^{\rm eq}\,,
\end{align}
where
\begin{align}
D=&[YY^\dagger]_{11} [YY^\dagger]_{22} (M_1^2-M_2^2)^2
\\\notag
+&\bar M^4\bar\Gamma^2([YY^\dagger]_{11} +[YY^\dagger]_{22})^2([YY^\dagger]_{11}[YY^\dagger]_{22}-{\rm Re}\{[YY^\dagger]_{12}\}^2)\,.
\end{align}
To obtain simple analytic results, we have specialised here on a case when only two sterile
neutrinos are dynamically relevant. For three and more sterile neutrinos in the game,
one may still
approximate Eq.~(\ref{nosc:cosmic}) by an algebraic equation when neglecting derivatives,
but one does not find closed forms for the solutions as simple as in a situation
that can be described by two sterile flavours only.

Comparing with Eqs.~(\ref{source:fl:NR}) and~(\ref{BEq:ell}), we
identify the time-dependent effective decay-asymmetry
\begin{align}
\label{epsilon:effective}
\varepsilon_{ab}(z)=
\frac{1}{16\pi}\frac{a_{\rm R}z}{\bar M}
\sum\limits_{\overset{i,j}{i\not=j}}
Y^*_{ia}Y_{jb}\left(\delta n^+_{0hij}-\delta n^-_{0h ij}\right)
\left(\frac{d}{dz}n^{\rm eq}\right)^{-1}\,.
\end{align}
It can be straightforwardly interpreted as the asymmetry yield per
sterile neutrino that drops out of equilibrium. This quantity differs from
the $CP$-violating parameter defined in Ref.~\cite{Iso:2014afa}, that
quantifies the yield in terms of the out-of-equilibrium neutrinos that are present
at a given point in time. The discrepancy is due to the time delay in the transition from diagonal
out-of equilibrium densities to off-diagonal correlations due to oscillations.
We write the late-time limit of the decay asymmetry~(\ref{epsilon:effective})
by dropping the argument $z$, {\it. e.g.}
$\varepsilon\equiv\varepsilon(\infty)$, for which we find when using Eq.~(\ref{deltan:latetime})
\begin{align}
\label{epsilon:flavoured}
\varepsilon_{ab}=&\frac{\bar M \bar\Gamma}{D}(M_1^2-M_2^2)\bar M  \left([YY^\dagger]_{11}+[YY^\dagger]_{22}\right){\cal Y}_{ab}\,,
\end{align}
where
\begin{align}
\label{Yukawa:ass}
{\cal Y}_{ab}=-\frac{\rm i}{2}\left(Y^\dagger_{a1}[YY^\dagger]_{12}Y_{2b}-Y^\dagger_{a2}[YY^\dagger]_{21}Y_{1b}+Y^\dagger_{a1}[Y^*Y^t]_{12}Y_{2b}-Y^\dagger_{a2}[Y^*Y^t]_{21}Y_{1b}
\right)
\,.
\end{align}
Provided the strong washout approximation holds, it is then easy to solve Eq.~(\ref{nosc:cosmic})
numerically. In the fully flavoured regime, $q_{\ell ab}$ can be reduced to its diagonal components
and the flavoured asymmetry can be calculated in straightforward generalisation (see {\it e.g.} Refs.~\cite{Blanchet:2006be,Garbrecht:2012pq})
of the methods for the single-flavour case~\cite{Kolb:1983ni,Buchmuller:2004nz}.

The flavoured expression~(\ref{epsilon:flavoured})
for the decay asymmetry in resonant Leptogenesis is
of importance throughout the parameter space. If the sterile neutrino mass is below
$10^{9}\,{\rm GeV}$, the usual treatment of flavoured Leptogenesis should apply, {\it i.e.}
$\varepsilon_{ab}$ can be reduced to its diagonal components, because interactions mediated
by SM-lepton Yukawa-couplings effectively erase all coherence~\cite{Abada:2006fw,Nardi:2006fx}.
(See however Ref.~\cite{Dev:2014laa} for
a counterexample, where even Yukawa-suppressed correlations at low temperature are of importance,
due to a special flavour alignment.) At higher temperatures, when the asymmetry results from
the decay of one sterile neutrino only, it is sufficient to either deal with
two (a linear combination
of $e$ and $\mu$) or one single flavour (a linear combination
of $e$, $\mu$ and $\tau$) only. Once the decay of more than one neutrino contributes, as it
is the case for resonant Leptogenesis,
there will be decay asymmetries in different linear combinations~\cite{Blanchet:2011xq,Antusch:2010ms}
that in general cannot be aligned simultaneously.
It then appears simplest to take the full expression for $\varepsilon_{ab}$, including
the off-diagonal correlations, and compute their
evolution following Ref.~\cite{Beneke:2010dz} (see also Ref.~\cite{Dev:2014laa}).

\section{Applicability of Approximations}

The effective decay asymmetry~(\ref{epsilon:flavoured}) and the equation for the
evolution of the lepton asymmetry~(\ref{BEq:ell}) offer a simple way of accurately
calculating the freeze-out asymmetry even in the resonant regime, where
approximations based on the mass splitting of the sterile neutrinos
being larger than their width are not applicable. In
order to describe the parametric range of validity of neglecting derivatives acting on
$\delta n^\pm_{0h}$ in Eq.~(\ref{nosc:cosmic}) more precisely, we first take the simplifying assumption of a single lepton
flavour only. The effective decay asymmetry can then be expressed in the simple form
\beq
\label{eq:final_asym_single}
\varepsilon = \frac{X \mathrm{sin}(2\varphi)}{X^2 +  \mathrm{sin}^2(\varphi)}\,,
\eeq
where $X$ is a dimensionless parameter defined as
\beq
\label{def:X}
X = \frac{\Delta}{\bar \Gamma (y_1^2 + y_2^2)}\,,
\eeq
and where $\Delta=\frac{M_1^2-M_2^2}{\bar M^2}$ is
the normalised mass difference, $y_{1,2}=|Y_{1,2}|$ and $\varphi$ is the relative phase of the Yukawa couplings, $\varphi=\arg(Y_2/Y_1)$.
Note that the solutions to Eq.~(\ref{nosc:cosmic}) remain unaltered as a function of $z$,
provided we leave the ratios $\bar M:\Delta:Y^2$ invariant. Therefore,
such a rescaling leaves $\varepsilon(z)$ and the late-time solutions unchanged as
well. This invariance can also be explicitly observed in
the late-time asymmetry~(\ref{eq:final_asym_single}).

The late-time asymmetry~(\ref{eq:final_asym_single}) can also
be constructed from the solutions given in Ref.~\cite{Iso:2014afa}, such that
we note agreement with the results of that work. However, our definition for
$\varepsilon$ differs from the $CP$-violating parameter proposed in Ref.~\cite{Iso:2014afa}.
Our choice is motivated by the fact that the result~(\ref{eq:final_asym_single})
quantifies the yield of lepton asymmetry in a transparent manner and
that it allows for a straightforward calculation of the final asymmetry,
provided the late-time limit is a good approximation at the time of freeze out,
what we illustrate in the remainder of this Section.

The expression for the late-time decay asymmetry~(\ref{eq:final_asym_single}) only leads
to an accurate approximation for the process of Leptogenesis, provided
the solutions to Eq.~(\ref{nosc:cosmic}) reach their late-time form,
where the derivatives acting on $\delta n_{0h}^\pm$ may be neglected,
prior to the freeze-out of the asymmetry.
Based on this requirement,
we derive a more precise analytical condition that allows to identify the parametric regions
where neglecting the derivatives of $\delta n^\pm_{0h}$ is indeed justified.
Since $\delta n_{0h}^\pm$ are Hermitian two by two matrices and moreover,
$n_{0h}^+=n_{0h}^{-t}$, Eq.~(\ref{nosc:cosmic}) corresponds to a coupled set
of four real differential equations. The smallest eigenvalue\footnote{The eigenvalues presented
in this work are for notational simplicity understood as {\it minus one} times the actual eigenvalues
of Eq.~(\ref{nosc:cosmic}). The latter have negative real parts, because the equation describes the
relaxation of $\delta n^\pm_{0h}$ toward {\it zero}.}
in vicinity of
the parametric points where $\varepsilon$ is close to unity [{\it cf.} Eq.~(\ref{eps:opti})]
is given by $\epsilon=\epsilon_{{\rm R}2}$, which is presented explicitly
by Eq.~(\ref{epsilons:single:flavour}), or alternatively by
\begin{align}
\label{eq:smallesteigenval_y1y2}
\epsilon=&
\frac{a_{\rm R}z}{2\bar M}
\Bigg[
2\bar y^2\bar\Gamma-
\frac{1}{\sqrt2}
\bigg(
-\Delta^2+4\bar y^4\bar\Gamma^2-4 y_1^2y_2^2\bar\Gamma^2\sin^2\varphi
\\\notag
+&\Big[
\Delta^4+(4 \bar y^4 -4 y_1^2 y_2^2 \sin^2\varphi)^2\bar\Gamma^4
+2\Delta^2\bar\Gamma^2
\left(
4 \bar y^4+4y_1^2y_2^2(\sin^2\varphi-2)
\right)
\Big]^{\frac12}
\bigg)^{\frac12}
\Bigg]\,,
\end{align}
where $\bar y^2=(y_1^2+y_2^2)/2$.
Notice also that $\epsilon$ is invariant when keeping the ratio $\bar M:\Delta:Y^2$
fixed. This is more easily seen in the democratic case $y_1=y_2 $, where the smallest eigenvalue is given by
\beq
\frac{\epsilon}{\bar \epsilon} = 1 - \vartheta(\cos^2\varphi-X^2)\sqrt{\cos^2\varphi-X^2}
\label{eq:ratio_x_phi}
\,,
\eeq
where $\vartheta$ is the Heaviside step function and where we
have defined $\bar\epsilon= (a_{\rm R} z/\bar M)\bar y^2 \bar\Gamma$.
Since $(dn^{\rm eq}/dz)/n^{\rm eq}={\cal O}(1)$ around freeze out,
one should require $\epsilon\gg1$ in order to neglect derivatives acting on
$\delta n^\pm_{0h}$. [A condition that
amounts to requiring that the slowest eigenmode of Eq.~(\ref{nosc:cosmic}) is faster
than the Hubble expansion rate.] This also implies that $\bar\epsilon\gg\bar\epsilon/\epsilon$.
The quantity $\bar\epsilon/\epsilon$ therefore is of phenomenological interest,
because it indicates how strong the washout must at least be such that we can
justify the neglect of the derivatives of $\delta n_{0h}^\pm$. In order to relate
to the parameters that are typically employed in calculations on Leptogenesis, note that
$\bar\epsilon/z=\bar K=(K_1+K_2)/2$, where the $K_i=y_i^2\bar M\bar\Gamma/H|_{T=\bar M}$
are the usual washout
parameters~\cite{Buchmuller:2004nz}.
In order to satisfy $\epsilon\gg1$ at the time of freeze-out, that occurs
for $z=z_{\rm f}={\cal O}(10)$, it follows that we must require
\begin{align}
\label{validity:condition}
\bar K\gg(1/z_{\rm f})(\bar\epsilon/\epsilon)\,.
\end{align}
We can therefore use the ratio
$\bar\epsilon/\epsilon$ in order to infer the minimal washout strength that is necessary
for consistently neglecting the derivatives of $\delta n^\pm_{0h}$. 

Note that the washout strength $\bar K$ can also be employed as an expansion parameter for
a series approximation that
generalises the truncation of the derivative of $\delta n_{0h}^\pm$
in Eq.~(\ref{nosc:cosmic}) in a systematic manner. Details of this are worked out
in Appendix~\ref{appendix:mixosc:expansion}.

%


It is interesting to consider the situation where, for a given value of $X$,
the phase $\varphi$ maximises the
decay asymmetry~(\ref{eq:final_asym_single}). This occurs for $\varphi=\varphi_M$,
where
\begin{align}
\label{phi:opti}
\varphi_M=\arctan\frac{X}{\sqrt{1 + X^2}}\,,
\end{align}
and where the asymmetry is then given by 
\begin{align}
\label{eps:opti}
\varepsilon=\frac{1}{\sqrt{1+X^2}}\,.
\end{align}
For $X\to 0$, the decay asymmetry attains its maximum value
$\varepsilon\to 1$. Curiously, in this case the $CP$-violating
phase tends to be vanishing, $\varphi_M\to0$.
The exact limit can however not be reached because for such an alignment
scenario, it takes infinitely long for the off-diagonal correlations in $\delta n^\pm_{0h}$
to build up. In particular, this does not occur before freeze-out. In the examples below, we
observe however that it is possible in practice to obtain asymmetries that are at least
close to maximal.

For comparison,
we also comment  the opposite regime, where $X\gg1$ (which may still allow for $\Delta\ll 1$). In that case the asymmetry is maximal when $\varphi_M(X\gg1) = \pi/4$.

\begin{figure}[t!]
\begin{center}
\epsfig{file=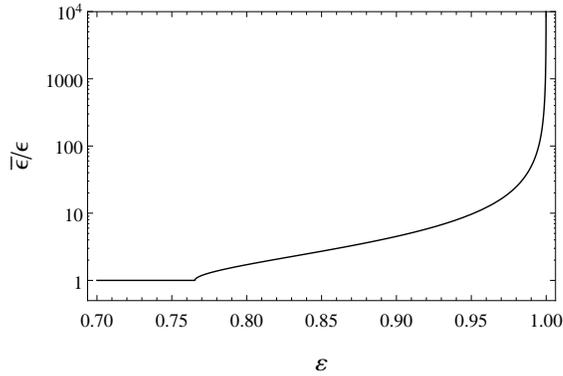,width=7.5cm}
\end{center}
\caption{
\label{fig:EVratio:phi:opti}
The ratio $\bar\epsilon/\epsilon$
of the diagonal relaxation rate of the sterile neutrinos to the smallest eigenvalue, with
$\varphi$ given by Eq.~(\ref{phi:opti}). In order
for the derivatives of $\delta n_{0h}$ to be negligible, the washout
strength should satisfy relation~(\ref{validity:condition}).
}
\end{figure}

\begin{figure}[t!]
\begin{center}
\epsfig{file=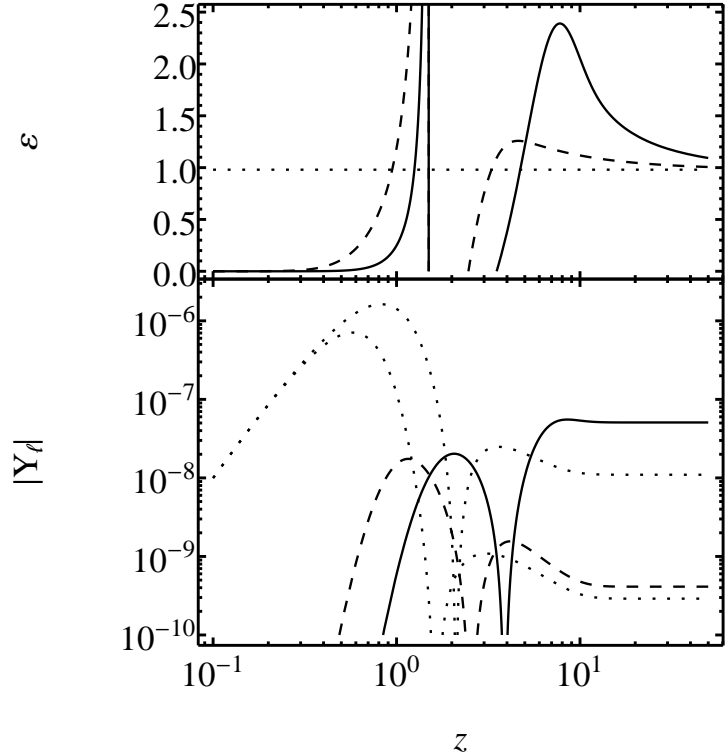,width=10.5cm}
\end{center}
\caption{
\label{fig:example:phi:opti}
Upper panel:
Evolution of the parameter $\varepsilon(z)$ toward the late-time limit
$\varepsilon=0.98$ (dotted) and the value~(\ref{phi:opti})
for $\varphi$ that maximises the asymmetry. We choose two different washout strengths,
$\bar K=5$ (solid) and $\bar K=20$ (dashed).
Lower panel:
Lepton asymmetry $|Y_\ell|=|\Delta_\ell|/s$ obtained from Eq.~(\ref{BEq:ell}) and with the
single-flavour simplifications
explained in the text, obtained with the time-dependent solution for
$\varepsilon(z)$ and $\bar K=5$ (solid) and $\bar K=20$ (dashed) and with
with the late-time limit $\varepsilon=0.98$ (dotted) (the cases
$\bar K=5$ and $\bar K=20$ are distinguishable by their proximity to
the solutions for $z$-dependent $\varepsilon(z)$).
}
\end{figure}

Substituting $\varphi=\varphi_M$ and the value of $X^2$ in terms of $\varepsilon$
from relation~(\ref{eps:opti}) into
Eq.~(\ref{eq:ratio_x_phi}), we find
\begin{align}
\label{eps:epsbar:vareps}
\frac{\epsilon}{\bar\epsilon}=1-\vartheta(\varepsilon^2-2+\sqrt{2})\sqrt{\frac{-(\varepsilon^2-2-\sqrt{2})(\varepsilon^2-2+\sqrt{2})}{\varepsilon^2(2-\varepsilon^2)}}
\,.
\end{align}
This ratio vanishes  as the asymmetry $\varepsilon$ goes to $ 1$,
which reflects the fact that for large asymmetries, it takes a longer time to build the
off-diagonal correlations in $\delta n^\pm_{0h}$, and the washout should be sufficiently
strong in order for the late-time decay asymmetry $\varepsilon$ to be a good approximation.
The ratio $\bar\epsilon/\epsilon$ is presented in
Figure~\ref{fig:EVratio:phi:opti}.

As an illustration for how to interpret 
the quantity $\bar\epsilon/\epsilon$, in Figure~\ref{fig:example:phi:opti}, we show
how the parameter $\varepsilon(z)$ [as defined in Eq.~(\ref{epsilon:effective})]
evolves in the case where it approaches
the late-time value $\varepsilon=0.98$.
We choose two washout strengths, where the weaker one violates the
criterion~(\ref{validity:condition}) while the stronger one marginally complies
with it. In order to obtain these results, we
assume vanishing initial distributions for the sterile neutrinos and begin to integrate
at $z=0$. We observe indeed that when relation~(\ref{validity:condition}) holds,
where $z_{\rm f}={\cal O}(10)$, a stationary form for $\varepsilon(z)$
corresponds to a good approximation.
To see the effect on the freeze-out lepton asymmetry, we take both, the late-time
value $\varepsilon$ and the time-dependent solution $\varepsilon(z)$, and solve
Eq.~(\ref{BEq:ell}), where we assume one single flavour
(and consequently suppress the flavour indices), set
$q_\phi=0$ for simplicity and
take $q_\ell=-\Delta_\ell$. We express the result in terms of the
ratio of the lepton-number to the entropy density $s$, $Y_\ell=-\Delta_\ell/s$
and use the value for $s$ with $106.75$ relativistic degrees of freedom.
For both washout strengths, we observe that initially, there is a substantial deviation between the
solutions for $Y_\ell$ that are based on the time dependent $\varepsilon(z)$ and its late-time
limit. While for the larger washout strength, the freeze-out asymmetries agree eventually up to
about $40\%$ accuracy, there is a discrepancy of about a factor of five for the smaller
washout strength, that does clearly not satisfy relation~(\ref{validity:condition}).

\begin{figure}[t!]
\begin{center}
\epsfig{file=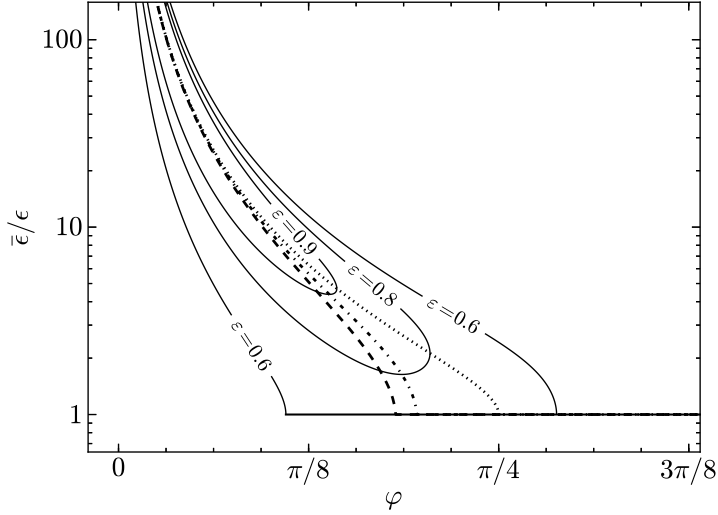,width=10.5cm}
\end{center}
\caption{
\label{fig:EVratio:phi:fixed}
Ratio $\bar \epsilon/\epsilon$ for several different late-time asymmetries.
Since Eq.~(\ref{eq:final_asym_single}) has two solutions for $X$, there also are two possible eigenvalues, given $\varphi$ and $\varepsilon$. The dotted line indicates the border between the two solutions. 
The ratio for $\varphi=\varphi_M$ from Eq.~(\ref{rel:EVs:phiM}) is shown with a dashed line, while the minimal eigenvalue ratio Eq.~(\ref{rel:EVs:minimum}) for a given $\varepsilon$ is given by the dot-dashed one.
Only the interval $[0,\pi/2]$ is shown here, because $\epsilon(\varphi)=\epsilon(\pi-\varphi)$.
}
\end{figure}

\begin{figure}[t!]
\begin{center}
\epsfig{file=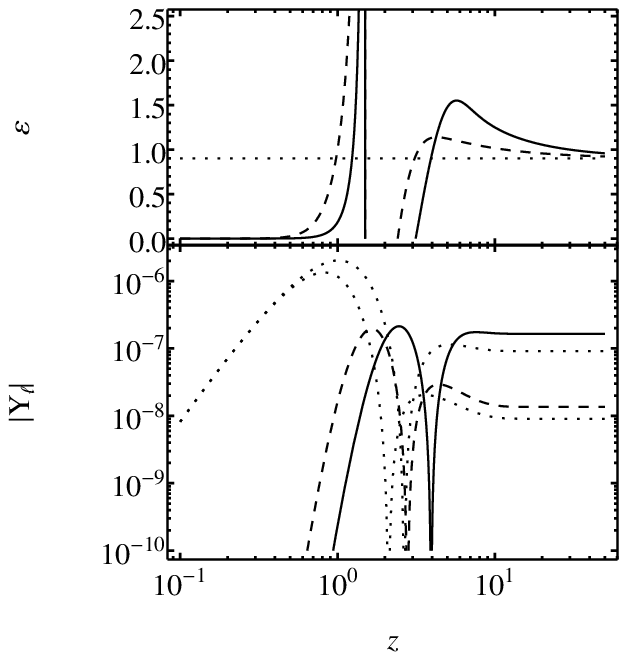,width=10.5cm}
\end{center}
\caption{
\label{fig:example:phifixed}
Upper panel:
Evolution of the parameter $\varepsilon(z)$ toward the late-time limit
$\varepsilon=0.9$ (dotted). We choose $\varphi=0.4$ and two different washout strengths,
$\bar K=2$ (solid) and $\bar K=5$ (dashed).
Lower panel:
Lepton asymmetry $|Y_\ell|=|\Delta_\ell|/s$ obtained from Eq.~(\ref{BEq:ell}) and with the
single-flavour simplifications
explained in the text, obtained with the time-dependent solution for
$\varepsilon(z)$ and $\bar K=2$ (solid) and $\bar K=5$ (dashed) and
with the late-time limit $\varepsilon=0.98$ (dotted) (the cases
$\bar K=2$ and $\bar K=5$ are distinguishable by their proximity to
the solutions for $z$-dependent $\varepsilon(z)$).
}
\end{figure}

Next, we again take $y_1=y_2$ but impose fixed values
of $\varphi$, in order to allow for a deviation from the relation~(\ref{phi:opti}).
In Figure~\ref{fig:EVratio:phi:fixed},
the ratios $\bar\epsilon/\epsilon$ are presented as functions
of $\varphi$ for various values of $\varepsilon$.
The curves exhibit two branches, because for
a given asymmetry $\varepsilon$ and phase $\varphi$, Eq.~(\ref{eq:final_asym_single})
has two solutions for $X$. The two branches join at the point where there is only one root.
It is easy to show, using Eq.~(\ref{eq:final_asym_single}), that the condition for a unique root is $\varepsilon = \mathrm{cos}(\varphi)$, for which $X = \mathrm{sin}(\varphi)$.
There are two more curves that we display in Figure~\ref{fig:EVratio:phi:fixed}. First, we show
the ratios of the eigenvalues when identifying $\varphi=\varphi_M$, what fixes $X$
through Eqs.~(\ref{phi:opti}),
and with Eq.~(\ref{eps:epsbar:vareps}), we obtain
\begin{align}
\label{rel:EVs:phiM}
\frac{\epsilon}{\bar\epsilon}=1-\vartheta\left( \mathrm{cos}^2(\varphi_M) - \frac{\mathrm{tan}^2 \varphi_M}{1 - \mathrm{tan}^2 \varphi_M}\right)\sqrt{ \mathrm{cos}^2(\varphi_M) - \frac{\mathrm{tan}^2 \varphi_M}{1 - \mathrm{tan}^2 \varphi_M}}
\,.
\end{align}
Second, we determine the value of $\varphi$ that minimises the eigenvalue ratio, what
defines the graph
\begin{align}
\label{rel:EVs:minimum}
\frac{\epsilon}{\bar\epsilon} = 1-\cos (\varphi ) \sqrt{1-\sin ^2(\varphi ) \sec (2 \varphi )}\,.
\end{align}
From Figure~\ref{fig:EVratio:phi:fixed}, we observe asymptotic proximity between
these two curves~(\ref{rel:EVs:phiM})
and~(\ref{rel:EVs:minimum}), and moreover, one
can check that the junction points for the two solutions for $X$ are close to these curves
as well.
This implies that
$\varphi=\varphi_M$ corresponds to a preferable choice for obtaining large asymmetries not only
because it maximises $\varepsilon$ but also because at the same time, it
minimises $\bar \epsilon/\epsilon$ and therefore
the required washout strength.

Again, we present
in Figure~\ref{fig:example:phifixed} the
evolution of the parameter $\varepsilon(z)$ and the lepton-number to entropy ratio
$Y_\ell$
for two different washout strengths, what exemplifies the use of the 
criterion~(\ref{validity:condition}) for approximating the freeze-out asymmetry using the late-time decay asymmetry $\varepsilon$. 

We now move from the simplifying single-flavour model to a more realistic
scenario, where several flavours are present and where
we take account of constraints from neutrino oscillation data. In order
to avoid a proliferation of free parameters, we consider the case where there
are only two sterile neutrinos or, alternatively, where a third sterile neutrino decouples.
It follows that
one of the masses $m_{1,2,3}$ of the
observed light neutrino states vanishes, {\it i.e.} $m_1=0$ for a normal
mass hierarchy, which is what we assume here. This leads to
a simplified form of the Casas-Ibarra parametrisation of the Yukawa couplings~\cite{Casas:2001sr}
\begin{align}
Y^\dagger=\frac{\sqrt 2}{v}U_\nu
\left(
\begin{array}{cc}
0 & 0\\
\sqrt{m2} & 0\\
0 & \sqrt{m3}
\end{array}
\right)
\left(
\begin{array}{cc}
-\sin\varrho & \cos\varrho\\
-\cos\varrho & -\sin\varrho
\end{array}
\right)
\left(
\begin{array}{cc}
\sqrt{M1} & 0\\
0 & \sqrt{M2}
\end{array}
\right)
\,,
\end{align}
where $U_\nu$ is the PMNS matrix and $v=246\,{\rm GeV}$ is the vacuum expectation value
of the Higgs field.
Note that here, $Y$ is a $2\times 3$ matrix. For the PMNS matrix
and for the light neutrino masses, we take the best-fit parameters from the global
analysis of Ref.~\cite{Fogli:2012ua} (see also~\cite{Forero:2014bxa}), and for simplicity, we fix the Dirac and
the Majorana phase therein to be zero. The parameter $\varrho$ is a complex angle, and
its imaginary part acts here in absence of the PMNS phases as the only
source of $CP$-violation. Moreover, this imaginary part largely controls the absolute value
of $\cos\varrho$ and $\sin\varrho$, {\it i.e.} large imaginary parts imply a
large washout strength.

For definiteness, we are considering this setup at temperatures of about $10^8\,{\rm GeV}$, where
all second-generation but none of the first-generation
Yukawa couplings are in equilibrium. The qualitative picture does not change when going to different
temperatures, where other spectator fields give rise to ${\cal O}(10\%)$ corrections to
the freeze-out asymmetries~\cite{Barbieri:1999ma,Buchmuller:2001sr,Davidson:2008bu}.
We can then relate
\begin{align}
\label{eq:spectators}
q_\ell=A \Delta_\ell\;, q_\phi=C_\phi \Delta_\ell\,,
\end{align}
where
\begin{subequations}
\begin{align}
A=&\frac{1}{1074}
\left(
\begin{array}{ccc}
-906 & 120 & 120 \\
75 & -688 & 28 \\
75 & 28 & -688
\end{array}
\right)\,,\\
C_\phi=&
-\frac{1}{179}
\left(
\begin{array}{ccc}
37 & 52 & 52
\end{array}
\right)\,.
\end{align}
\end{subequations}
Moreover, at temperatures below $10^9\,{\rm GeV}$, the off-diagonal correlations of the
left-handed leptons are strongly suppressed due to the SM Yukawa interactions, such that
we can neglect the off-diagonal elements of Eq.~(\ref{BEq:ell}) (see however Ref.~\cite{Dev:2014laa},
where due to alignments of the Yukawa couplings $Y$ the off-diagonal correlations remain
non-negligible at even smaller temperatures).

\begin{figure}[t!]
\begin{center}
\epsfig{file=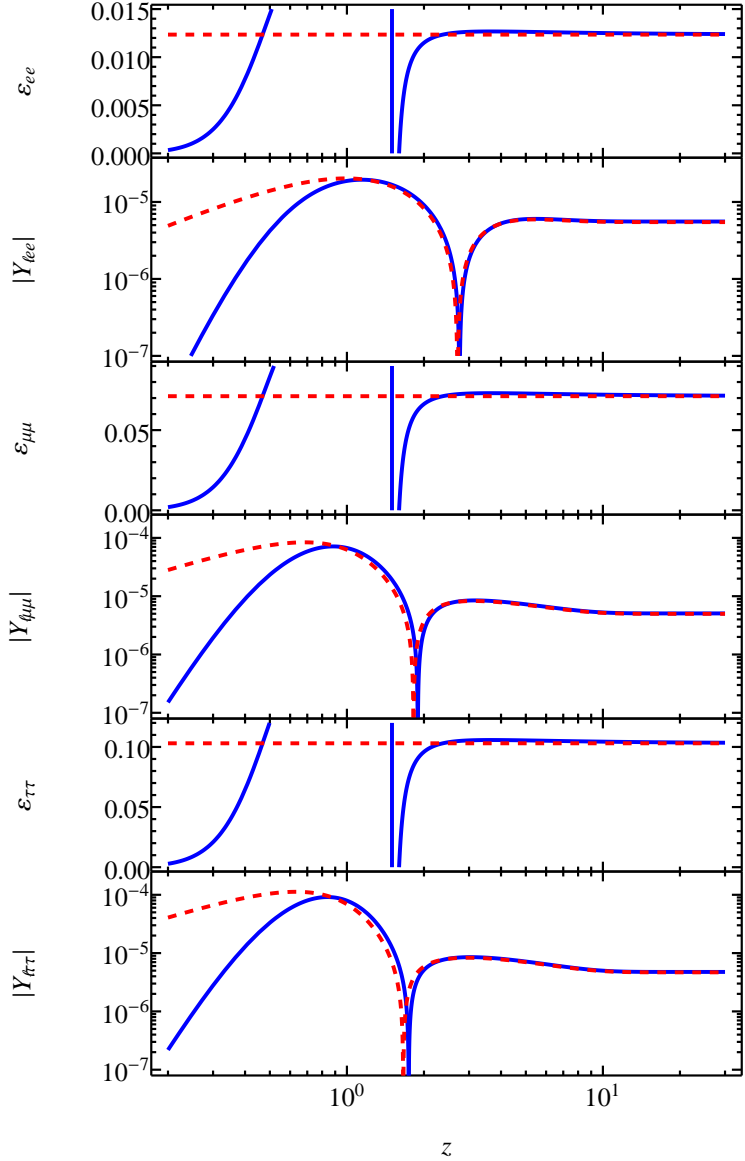,width=10.5cm}
\end{center}
\caption{
\label{fig:flavoured:models}
Time-dependent flavoured decay asymmetries from Eq.~(\ref{epsilon:effective}) (solid) compared
to their late-time limit $\varepsilon_{aa}$ from Eq.~(\ref{epsilon:flavoured}) (dashed).
We take the parameters $\delta=0$, $\alpha=0$, $\varrho=\pi/4+0.2{\rm i}$,
$\Delta/\bar M=-2\times 10^{-17}{\rm GeV}^{-1}$.
We also present the
individual flavoured baryon-minus lepton asymmetries $|Y_{\ell aa}|=|\Delta_{\ell aa}|/s$
obtained from Eq.~(\ref{BEq:ell}), using the time-dependent decay asymmetry (solid)
and the late-time limit (dashed). The quantities
$\Delta_{\ell aa}$, $q_{\ell aa}$ and $q_\phi$
are related through Eqs.~(\ref{eq:spectators}).
}
\end{figure}

It is also interesting to discuss the
radiative processes that lead to
small corrections to the leading-order rates for the production of the sterile neutrinos and the
washout of the lepton asymmetry that we employ in Eqs.~(\ref{nosc:cosmic})
and~(\ref{BEq:ell}). The dominating corrections are due to the
radiation of gauge bosons and top-quark Yukawa-interactions.
In the context of resonant Leptogenesis, these
are discussed in Ref.~\cite{Pilaftsis:2003gt}. There has been some recent progress
in that the cancellation of soft and collinear divergences in the thermal backround
was shown for non-relativistic sterile neutrinos, leading to a consistent calculation
of these rates in the strong washout regime~\cite{Salvio:2011sf,Laine:2011pq,Biondini:2013xua}.
The corrections are found to be at the
few percent level~\cite{Salvio:2011sf}, and therefore we do not include these in the calculations for
the present numerical examples. We note also that the cancellation of soft and collinear
divergences has recently been demonstrated as well for relativistic massive sterile
neutrinos in Refs.~\cite{Garbrecht:2013gd,Laine:2013lka}.
Besides, it was pointed out in Ref.~\cite{Bodeker:2014hqa}, that radiative corrections
(in particular the thermal masses) have a subleading effect on the spectator processes
because they change the
susceptibility relation between the chemical potentials and the charge densities.

The eigenvalues of the equation for mixing and oscillating sterile neutrinos~(\ref{nosc:cosmic})
in terms of the Casas-Ibarra parametrisation
are given in Eq.~(\ref{epsilons:multi:flavour}). As the oscillatory contributions due to the
mass splitting enter as an imaginary part and the damping contributions due to the Yukawa couplings
as a real part, we can find a lower bound on the magnitude of these eigenvalues by setting $\Delta=0$,
what leads to a considerable simplification of the expressions:
\begin{align}
\label{eps:CI:onresonance}
\epsilon^{\rm CI}_{{\rm I}1,2}/\bar \epsilon^{\rm CI}=1\,,\quad
\epsilon^{\rm CI}_{{\rm R}1,2}/\bar \epsilon^{\rm CI}=\frac{m_2+m_3\pm(m_3-m_2){\rm sech}(2{\rm Im}[\varrho])}{(m_2+m_3)}\,,\;
\end{align}
Since the smallest ratio is $\epsilon^{\rm CI}_{{\rm R}2}/\bar \epsilon^{\rm CI}\gsim 1/6$
for normal hierarchy,
neglecting the derivatives on $\delta n^{\pm}_{0h}$ in Eq.~(\ref{nosc:cosmic}) is
by the  criterion~(\ref{validity:condition}) (assuming $z_{\rm f}={\cal O}(10)$)
a good
approximation everywhere in the strong washout regime of resonant Leptogenesis
for the phenomenological model with
two sterile neutrinos. Moreover, as
washout is always strong in that scenario, what we show in Appendix~\ref{appendix:washout:2RHN},
we can conclude
that using the late-time asymmetry~(\ref{epsilon:flavoured}) is a valid approximation
for any point in parameter space.

For the phenomenological model specified above,
we solve Eq.~(\ref{BEq:ell}) with the
effective decay asymmetry~(\ref{epsilon:effective}) based on the full numerical
solution to Eqs.~(\ref{nosc:cosmic}). This, we compare with the solution
obtained when using the late-time limit for the decay
asymmetry~(\ref{epsilon:flavoured}) for all times prior to freeze-out.
Since by above arguments, 
there should be no points where the freeze-out asymmetries
obtained by the two methods differ by substantial amounts, 
we show in Figure~\ref{fig:flavoured:models} the evolutions
of $\varepsilon_{aa}(z)$ from Eq.~(\ref{epsilon:effective}) and the
values of $\varepsilon_{aa}$ from Eq.~(\ref{epsilon:flavoured}), along with
the asymmetries $|Y_{\ell aa}|=|\Delta_{\ell aa}|/s$ obtained using
the time-dependent and the effective late-time decay asymmetries for a 
typical point in parameter space, for which the width dominates the
mass splitting, $\Delta\ll({\rm tr}[Y Y^\dagger]\bar \Gamma)^2$. As anticipated from the analysis of
the eigenvalues, albeit the different time evolution at early stages,
the freeze-out asymmetries
agree very well.

\section{Comparison with other Regulators for the Decay Asymmetry}

Due to the conceptual interest in the
question of the behaviour of the decay asymmetry in the resonant regime,
a number of terms have been suggested earlier to avoid a
resonance catastrophe from the enhancement factor $1/(M_i^2-M_j^2)$ in the
degenerate limit $M_i\to M_j$. For the purpose of comparing with these results,
we recast the asymmetry~(\ref{epsilon:flavoured}) to the form
\begin{align}
\label{epsilon:recast}
\varepsilon_{ab}=&
\frac{{\cal Y}_{ab}}{8\pi}\left(\frac{1}{[YY^\dagger]_{11}}+\frac{1}{[YY^\dagger]_{22}}\right)\frac{\bar M^2 (M_1^2-M_2^2)}{(M_1^2-M_2^2)^2+R}\,,
\end{align}
where
\begin{align}
\label{R:strongwashout}
R=\frac{\bar M^4}{64\pi^2}\frac{([YY^\dagger]_{11}+[YY^\dagger]_{22})^2}{[YY^\dagger]_{11}[YY^\dagger]_{22}}\left(({\rm Im}[YY^\dagger]_{12})^2+\det YY^\dagger\right)\,.
\end{align}
Moreover, while the results of Refs.~\cite{Pilaftsis:2003gt,Anisimov:2005hr,Garny:2011hg} do not include active lepton flavour effects, it is easy
to supplement these with the flavour structure of the SM leptons, which we do here
for the sake of clarity of the comparison.

We should emphasise once more that the decay asymmetry~(\ref{epsilon:flavoured})
[or its equivalent form~(\ref{epsilon:recast}) with the regulator~(\ref{R:strongwashout})]
applies only to the strong washout regime and when all damping rates
in the linear differential equation~(\ref{nosc:cosmic}) are large compared
to the Hubble rate at the time of the freeze out of the asymmetry. In general,
the decay asymmetries will depend on how the initial state in terms of the sterile
neutrinos is prepared [{\it I.e.} for the formulae~(\ref{epsilon:flavoured},\ref{epsilon:recast}), the out-of-equilibrium neutrinos appear due to the expansion
of the Universe.], and there may be a time dependence, matters which
should be familiar from systems of mixing neutral mesons~\cite{Beringer:1900zz}.
However, the results of Refs.~\cite{Pilaftsis:2003gt,Anisimov:2005hr} were thought
to be universally applicable, which is not the case according to the
present work and other recent publications on resonant Leptogenesis~\cite{Garbrecht:2011aw,Garny:2011hg,Iso:2013lba,Iso:2014afa}.
We remark that the asymmetry from Ref.~\cite{Pilaftsis:2003gt} is supplemented
in Ref.~\cite{Dev:2014laa} by extra terms that describe the asymmetry from oscillations. While
it would be interesting to compare both approaches in detail, one may find the
path taken here, {\it i.e.} to attribute the entire asymmetry to
oscillations of sterile neutrinos as described by Eq.~(\ref{nosc:cosmic}), more economical.

We now quote some of the most widely discussed previous expressions for
the decay asymmetry for resonant Leptogenesis.
In Ref.~\cite{Pilaftsis:2003gt}, a regulator is obtained in the standard
$S$-matrix formalism by using
a resummed form for the intermediate propagator of the sterile neutrino
that occurs in the wave-function diagram, such that the sum of the decay asymmetries of two sterile neutrinos
is found to be
\begin{align}
\varepsilon_{ab}=&
\frac{{\cal Y}_{ab}}{8\pi}
\left(\frac{\bar M^2 (M_1^2-M_2^2)}{(M_1^2-M_2^2)^2+\frac{[YY^\dagger]_{22}^2}{64\pi^2}\bar M^4}\frac{1}{[YY^\dagger]_{11}}
+\frac{\bar M^2 (M_1^2-M_2^2)}{(M_1^2-M_2^2)^2+\frac{[YY^\dagger]_{11}^2}{64\pi^2}\bar M^4}\frac{1}{[YY^\dagger]_{22}}\right)\,.
\end{align}

Subsequently, in Ref.~\cite{Anisimov:2005hr} it is argued that the resummation needs to
take account of the mixing of both sterile neutrinos. This results in an expression of
the form~(\ref{epsilon:recast}) (provided we approximate $M_1 M_2\approx \bar M^2$),
but with the regulator term
\begin{align}
\label{regulator:Broncano}
R=\frac{\bar M^4}{64\pi^2}\left([YY^\dagger]_{11}-[YY^\dagger]_{22}\right)^2
\,.
\end{align}
When solving the equations for the oscillating sterile neutrinos, one may recover
the corresponding corrections for vanishing inital disributions for the sterile
neutrinos that relax towards thermal equilibrium with leptons and Higgs
particles~\cite{Garbrecht:2011aw,Garny:2011hg} (a setup tyically not applicable to cosmological contexts), provided the mass separation
is larger than the width of the sterile neutrinos. In the interesting degenerate
regime (mass splitting smaller than the width),
the regulator~(\ref{regulator:Broncano}) is however not applicable.

A correct form for the regulator for vanishing initial distributions
of almost mass-degenerate sterile neutrinos that relax to equilibrium is obtained instead in
Ref.~\cite{Garny:2011hg}:
\begin{align}
R=\frac{\bar M^4}{64\pi^2}\left([YY^\dagger]_{11}+[YY^\dagger]_{22}\right)^2
\,.
\end{align}
This analytic result relies on the assumption that
$|(YY^\dagger)_{12}|\ll|(YY^\dagger)_{11,22}|$ (what necessarily requires
the sterile neutrinos coupling to several flavours of active leptons),
under which the result~(\ref{R:strongwashout}) of this work reduces to the same form\footnote{We would like to thank M.~Garny for pointing this out.}.
This observation may be explained by the fact that provided $|(YY^\dagger)_{12}|\ll|(YY^\dagger)_{11,22}|$, the time-derivatives acting on the off-diagonal correlations
of the sterile neutrinos in the equations that describe the neutrino
oscillations are negligible.

\section{Conclusions}

We have studied the applicability of the late-time decay asymmetries $\varepsilon$
for sterile neutrinos
in their multi-flavoured and single-flavoured
forms~(\ref{epsilon:flavoured}) and~(\ref{eq:final_asym_single}) to computations of
the freeze-out asymmetry in resonant Leptogenesis. This has been done
by comparison with the results obtained from the time-dependent decay
asymmetry~(\ref{epsilon:effective}) that is based on the solution to
the evolution equation~(\ref{nosc:cosmic}) for the mixing and oscillating sterile neutrinos.
The evolution equation can be straightforwardly derived from its relativistic generalisation,
that was first presented in Ref.~\cite{Garbrecht:2011aw}. Following Ref.~\cite{Iso:2014afa},
the approximations~(\ref{epsilon:flavoured}) and~(\ref{eq:final_asym_single}) are obtained
by neglecting the time derivative acting on the non-equilibrium number densities
and correlations in Eq.~(\ref{nosc:cosmic}).

In addition to the numerical comparisons,
to gain analytical insight, we have derived expressions for the
eigenvalues of the equation that governs the mixing of the
sterile neutrinos and their deviation from equilibrium. This analysis reveals
that $\varepsilon$ can reach its maximum value {\it one} provided $\Delta\to 0$
and $\varphi\to 0$ simultaneously. In that case however, also the smallest eigenvalue
of the equation describing mixing and oscillations tends to {\it zero}, indicating
that the approximation in terms of the late-time decay asymmetry is not valid in that limit.
Nonetheless, the quantitative analysis (by studying the smallest eigenvalue as well
as the numerical solution) reveals that the late-time asymmetry can be a good approximation
already for moderately strong washout, even when $\varepsilon$ is close to {\it one}.
To quantify this, {\it cf.}
Figures~\ref{fig:EVratio:phi:opti} and~\ref{fig:EVratio:phi:fixed}
in conjunction with the criterion~(\ref{validity:condition}).
An increase of the washout strength generally
leads to a better approximation.

While the derivation
of the single-flavour decay asymmetry~(\ref{eq:final_asym_single}) makes use of
the approximation proposed in
Ref.~\cite{Iso:2014afa}, its definition is different from the
$CP$-violating parameter introduced in that work. We find the form that is suggested here
somewhat more transparent, as it corresponds to the asymmetry yield per sterile
neutrino that initially drops out of equilibrium through the Hubble expansion.
Moreover, with its definition as in the present work, the parameter $\varepsilon$
can be employed in the same way the usual vacuum decay asymmetry is used
in standard calculations on Leptogenesis~\cite{Giudice:2003jh,Buchmuller:2004nz,Davidson:2008bu}.
We have exemplified this point by explicitly calculating the freeze-out lepton asymmetry
in a phenomenological see-saw model that is consistent with the neutrino mixing and oscillation data.

We can draw the conclusion that the approximation proposed in Ref.~\cite{Iso:2014afa},
which leads to the late-time asymmetries that we derive and study here, is applicable
for Leptogenesis calculations in the strong washout regime of the single-flavour
model, unless the $CP$ asymmetry
and the mass splitting are very small simultaneously, {\it cf.} 
Eqs.~(\ref{eq:smallesteigenval_y1y2},\ref{eq:ratio_x_phi},\ref{epsilons:single:flavour})
and relation~(\ref{validity:condition}).
For the phenomenological model with two sterile neutrinos that is consistent with the
oscillations of active neutrinos, we find that the late-time asymmetries always
lead to a good approximation for the freeze-out values of the lepton number
densities. One potential caveat is that the early-time evolution of $\varepsilon(z)$
may strongly affect the asymmetry present within spectator fields, that in turn can
have a substantial impact on the freeze-out lepton asymmetry~\cite{Garbrecht:2014kda}. It should also be noted,
while the strong washout approximation always applies for resonant Leptogenesis with
two sterile neutrinos, this does not need not to be the case when more of these are present.
When the use of the late-time decay asymmetry cannot be justified,
one should simply replace it with the time-dependent decay
asymmetry~(\ref{epsilon:effective}) that is based on numerical solutions for the
mixing and the oscillations of the sterile neutrinos. Methods for obtaining accurate quantitative
results for Leptogenesis in the strong washout regime are therefore available throughout
parameter space.

\subsection*{Acknowledgements}
We would like to thank M.~Garny and A.~Kartavtsev for valuable comments on the
manuscript.
This work is supported in parts by the Gottfried Wilhelm Leibniz programme
of the Deutsche Forschungsgemeinschaft (DFG), by
a DFG Research Grant, by the DFG cluster of excellence `Origin and Structure of the Universe'
and by the National Science Foundation under Grant No. NSF PHY11-25915.
BG is grateful to the KITP at UC Santa Barbara for hospitality during completion of this work.

\begin{appendix}

\renewcommand{\theequation}{\Alph{section}\arabic{equation}}
\setcounter{equation}{0}

\section{Analytic Expansion of the Time Evolution}
\label{appendix:mixosc:expansion}
When the derivative of the equilibrium distribution is neglected, 
Eq.~(\ref{nosc:cosmic}) becomes homogeneous, and can be solved exactly:
\beq
\label{n:sol:hom}
\delta n_{0 h}^{\pm} (z) = {\rm e}^{(\mp \frac{{\rm i}}{2} \Omega -\frac{1}{2}\Gamma)\frac{z^2}{2} }\delta n_{0 h}^{\pm} (z=0) {\rm e}^{(\pm \frac{{\rm i}}{2} \Omega -\frac{1}{2}\Gamma)\frac{z^2}{2} }\,,
\eeq
where $\Omega$ is given by
\beq
\Omega =  \frac{a_{\rm R}}{\bar M} \frac{M^2}{\bar M^2} = \bar K \begin{pmatrix} X& 0\\ 0 & -X \end{pmatrix},
\eeq
and $\Gamma$ by
\beq
\Gamma = \frac{a_{\rm R}}{\bar M} \bar \Gamma {\rm Re}[Y^* Y^t] = \begin{pmatrix} K_1& \sqrt{K_1 K_2} \cos \varphi \\ \sqrt{K_1 K_2} \cos \varphi & K_2 \end{pmatrix}\,,
\eeq
where $\bar K=(K_1+K_2)/2$ and $X$ can in the single flavour case be identified with the parameter
defined in Eq.~(\ref{n:sol:hom}).

To obtain the solutions, a matrix $\Xi$ is defined, similarly to the one in Ref.~\cite{Garbrecht:2011aw}:
\beq
\Xi = (\Gamma + {\rm i} \Omega)/2\,.
\eeq
The solution~(\ref{n:sol:hom}) can now be rewritten as:
\beq
\begin{split}
\delta n_{0 h}^{+} (z) =& {\rm e}^{-\Xi \frac{z^2}{2} }\delta n_{0 h}^{+} (z=0) {\rm e}^{-\Xi^* \frac{z^2}{2} }\\
=& U^{-1} {\rm e}^{-\Xi_D \frac{z^2}{2}} U \delta n_{0 h}^{+} (z=0) V^{-1} {\rm e}^{-\Xi_D^* \frac{z^2}{2}} V
\,,
\end{split}
\eeq
where the matrices $U$ and $V$ diagonalise $\Xi$ and $\Xi^*$. The corresponding
eigenvalues $\Xi_D$ are:
\beq
\Xi_{D 1,2} = \frac{\bar K}{2} \left(1 \mp 
\sqrt{\cos ^2\varphi +{\Delta_K}^2 \sin ^2\varphi -X^2+2 {\rm i} \Delta_K X} \right)
\,,
\eeq
where $\Delta_K = (K_1-K_2)/(2 \bar K)$, which is zero in the democratic case.
We define $\gamma$ and $\omega$ as the real and imaginary parts of the above root.
\beq
\gamma + {\rm i} \omega = \sqrt{\cos ^2(\varphi )+{\Delta_K}^2 \sin ^2(\varphi )-X^2+2 {\rm i} \Delta_K X}\,.
\eeq
The transformation matrix $U$ is then given by:
\beq
U = c \begin{pmatrix} \sqrt{1-\Delta_K^2} \cos \varphi & -(\gamma + {\rm i} \omega +\Delta_K + {\rm i} X) \\ \gamma + {\rm i} \omega +\Delta_K + {\rm i} X & \sqrt{1-\Delta_K^2} \cos \varphi \end{pmatrix}
\,.
\eeq
In the case of a symmetric matrix $\Xi$, if $c$ is chosen such that $\det(U)=1$, the matrix inverse can be calculated as $U^{-1}=U^\mathrm{T}$, and there is also the relation $V=U^*$.
Rewriting Eq.~(\ref{nosc:cosmic}) in terms of $\Xi$ and $\Xi^*$, we can easily obtain the eigenmatrices:
\beq
\begin{split}
e_{11} = U^{-1}\begin{pmatrix} 1 & 0 \\ 0 & 0 \end{pmatrix} V\,, \qquad & e_{12} = U^{-1}\begin{pmatrix} 0 & 1 \\ 0 & 0 \end{pmatrix} V\,,\\
e_{21} = U^{-1}\begin{pmatrix} 0 & 0 \\ 1 & 0 \end{pmatrix} V\,, \qquad & e_{22} = U^{-1}\begin{pmatrix} 0 & 0 \\ 0 & 1 \end{pmatrix} V\,,
\end{split}
\eeq
and the corresponding eigenvalues:
\beq
\begin{split}
&\lambda_{11}=\Xi_1 + \Xi_1^* = \bar K (1-\gamma)\,,\\
&\lambda_{12}=\Xi_1 + \Xi_2^* = \bar K (1-i\omega)\,,\\
&\lambda_{21}=\Xi_2 + \Xi_1^* = \bar K (1+i\omega)\,,\\
&\lambda_{22}=\Xi_2 + \Xi_2^* = \bar K (1+\gamma)\,.
\end{split}
\eeq
It is important to notice here that $\lambda_{11}$ is equal to the smallest eigenvalue $\epsilon$ from Eq.~(\ref{eq:smallesteigenval_y1y2}) up to a factor of $z$.
\beq
\lambda_{11} = \frac{\epsilon}{z} = \bar K [1 - \vartheta(\cos^2\varphi-X^2)\sqrt{\cos^2\varphi-X^2} ]\,.
\eeq
Now that we have the eigenvalues and eigenmatrices for the homogeneous system, we can find a solution for the inhomogeneous case. It can be constructed as:
\beq
\label{n:sol:inhom}
\begin{split}
\delta n_{0 h}^{+} (z) &= -{\rm e}^{-\Xi \frac{z^2}{2} } \left[ \int^z  {\rm e}^{\Xi \frac{z'^2}{2} } \dfrac{\mathrm d n^{\rm eq}}{\mathrm d z'}  {\rm e}^{\Xi^* \frac{z'^2}{2} } \mathrm{d}z'\right] {\rm e}^{-\Xi^* \frac{z^2}{2} }\\
&=- U^{-1} {\rm e}^{-\Xi_D \frac{z^2}{2}}\left[
\int^z 	{\rm e}^{\Xi_D \frac{z'^2}{2}}U \dfrac{\mathrm d n^{\rm eq}}{\mathrm d z'}V^{-1} {\rm e}^{\Xi_D^* \frac{z'^2}{2}} \mathrm d z'
\right] {\rm e}^{-\Xi_D^* \frac{z^2}{2}} V\,.
\end{split}
\eeq
Next, we isolate the integral
\beq
\left[
\int^z 	{\rm e}^{\Xi_D \frac{z'^2}{2}}U \dfrac{\mathrm d n^{\rm eq}}{\mathrm d z'}V^{-1} {\rm e}^{\Xi_D^* \frac{z'^2}{2}} \mathrm d z'
\right]_{ij}
=
[U V^{-1}]_{ij} \int^z {\rm e}^{\lambda_{ij} z'^2/2} \dfrac{\mathrm d n^{\rm eq}}{\mathrm d z'} \mathrm d z'
\,,
\eeq
substitute $\tau = z'^2/2$ and then integrate by parts, what resuls in the series
\beq
\begin{split}
\int^{z^2/2} {\rm e}^{\lambda_{ij} \tau} \dfrac{\mathrm d n^{\rm eq}}{\mathrm d \tau} \mathrm d \tau =& \left. \lambda_{ij}^{-1}{\rm e}^{\lambda_{ij} \tau} \dfrac{\mathrm d n^{\rm eq}}{\mathrm d \tau}\right|^{z^2/2} - \int^{z^2/2} {\rm e}^{\lambda_{ij} \tau} \dfrac{\mathrm d^2 n^{\rm eq}}{\mathrm d \tau^2} \mathrm d \tau \\
=&{\rm e}^{\lambda_{ij} z^2/2} \left. \left[\frac{1}{\lambda_{ij}} \dfrac{\mathrm d n^{\rm eq}}{\mathrm d \tau} - \frac{1}{\lambda_{ij}^2} \dfrac{\mathrm d^2 n^{\rm eq}}{\mathrm d \tau^2} + \frac{1}{\lambda_{ij}^3} \dfrac{\mathrm d^3 n^{\rm eq}}{\mathrm d \tau^3} + \dots \right] \right|^{z^2/2}
\,.
\end{split}
\eeq
Using this result in the expression for the particular solution~(\ref{n:sol:inhom}), we obtain:
\begin{eqnarray}
\label{dltnKexpand}
	[\delta n^+_{0h}(z)]_{hk} &=& U^{-1}_{hi}(U V^{-1})_{ij}V_{jk} {\rm e}^{-\Xi_i z^2/2} {\rm e}^{\lambda_{ij} z^2/2} {\rm e}^{-\Xi_j^* z^2/2 } \sum_{m=1}^{\infty} \left(\frac{-1}{\lambda_{ij}}\right)^m \dfrac{\mathrm d^m n^{\rm eq}}{\mathrm d \tau^m}\\
&=& U^{-1}_{hi}(U V^{-1})_{ij}V_{jk}\sum_{m=1}^{\infty} \left(\frac{-1}{\lambda_{ij}}\right)^m \dfrac{\mathrm d^m n^{\rm eq}}{\mathrm d \tau^m}\,.
\end{eqnarray}

As only off-diagonal terms enter the source, we only need to calculate $\mathrm{Im} (\delta n^+_{0h})$:
\beq
\mathrm{Im}[\delta n^+_{0h,12}(z)]=\sum_{m=1}^{\infty} \zeta_m \left(\frac{-1}{\bar K}\right)^m \dfrac{\mathrm d^m n^{\rm eq}}{\mathrm d \tau^m}\,,,
\eeq
where we have introduced $\zeta_m$, which can be obtained
by multiplying the matrices in Eq.~(\ref{dltnKexpand}). In the democratic case,
it takes the form:
\beq
\zeta_n = -\frac{X \cos \varphi }{2 (\cos^2\varphi - X^2)} \left(2-\left(1-\gamma \right)^{-n}-\left(1+\gamma \right)^{-n}\right)
\eeq
We show the first few coefficients $\zeta_m$ in Table~\ref{table:zeta}.
\begin{table}
\begin{center}
\begin{tabular}{c|c}
$m$ & $\zeta_m$ \\ 
\hline 
1 & $\frac{X \cos \varphi }{\sin^2 \varphi + X^2}$ \\ 
\hline 
2 & $-\frac{X \cos (\varphi ) \left(\cos (2 \varphi )-2 X^2-5\right)}{2 \left( \sin^2 \varphi + X^2 \right)^2}$ \\ 
\hline
3 & $\frac{X \cos (\varphi ) \left(\cos (4 \varphi )-8 \left(X^2+1\right) \cos (2 \varphi )+8 \left(X^2+2\right) X^2+39\right)}{8 \left(\sin^2 \varphi + X^2 \right)^3}$ \\ 
\end{tabular}
\end{center}
\caption{\label{table:zeta} The first three coefficients $\zeta_m$}
\end{table}
When neglecting terms of order $\propto 1/\bar K^2$ and higher, one can easily obtain the late-time effective decay asymmetry~(\ref{eq:final_asym_single}).

\setcounter{equation}{0}
\section{Eigenvalues in the CI parametrization}

%
%
%
%
%

In the single flavour case, the eigenvalues of Eq.~(\ref{nosc:cosmic}) for the mixing
and oscillating sterile neutrinos are given by
\begin{align}
\label{epsilons:single:flavour}
\epsilon_{{\rm R}1,2}=\frac{a_{\rm R} z}{2\bar M}\left(
(y_1^2+y_2^2)\bar\Gamma\pm{\rm Re}[\sqrt D]
\right)
\quad
\textnormal{and}
\quad
\epsilon_{{\rm I}1,2}=\frac{a_{\rm R} z}{2\bar M}\left(
(y_1^2+y_2^2)\bar\Gamma\pm{\rm i}{\rm Im}[\sqrt D]
\right)\,,
\end{align}
where
\begin{align}
D=(y_1^2-y_2^2)^2\bar\Gamma^2+4y_1^2y_2^2\bar\Gamma^2\cos^2\varphi +2{\rm i}\Delta(y_1^2-y_2^2)\bar \Gamma-\Delta^2\,.
\end{align}
Similarly, with the Casas-Ibarra parametrisation of the phenomenological model, we obtain
the eigenvalues
\begin{subequations}
\label{epsilons:multi:flavour}
\begin{align}
\epsilon^{\rm CI}_{{\rm R}1,2}=&\frac{a_{\rm R} z}{4\bar M v^2}\left(
\bar\Gamma\bar M \left[(m_3-m_2)\Delta\cos(2{\rm Re}[\varrho])+4(m_2+m_3)\cosh(2{\rm Im}[\varrho])\right]\pm{\rm Re}[\sqrt{D_{\rm CI}}]
\right)\,,
\\
\epsilon^{\rm CI}_{{\rm I}1,2}=&\frac{a_{\rm R} z}{4\bar M v^2}\left(
\bar\Gamma\bar M \left[(m_3-m_2)\Delta\cos(2{\rm Re}[\varrho])+4(m_2+m_3)\cosh(2{\rm Im}[\varrho])\right]\pm{\rm i}{\rm Im}[\sqrt{D_{\rm CI}}]
\right)\,,
\end{align}
\end{subequations}
where
\begin{align}
D_{\rm CI}=&4\bar\Gamma^2\bar M^2\left(1-\frac{\Delta^2}{16}\right)(m_3-m_2)^2\sin^2(2{\rm Re}[\varrho])
\\\notag
-&
\left[\Delta\left(v^2+{\rm i}\frac{\bar M\bar\Gamma}{2}(m_2+m_3)\cosh(2{\rm Im}[\varrho])\right)+2{\rm i}\bar M \bar\Gamma(m_3-m_2)\cos(2{\rm Re}[\varrho])\right]^2\,.
\end{align}
In order to compare the magnitude of the individual
eigenvalues, we define in addition and in analogy
with the single-flavour model the parameter
\begin{align}
\label{epsbar:CI}
\bar\epsilon^{\rm CI}=\frac{a_{\rm R} z}{2 \bar M}{\rm tr}[YY^\dagger]\bar\Gamma
=\frac{a_{\rm R} z}{4 \bar M}\frac{\bar\Gamma\bar M}{v^2}
\left[4(m_2+m_3)\cosh(2{\rm Im}[\varrho])+(m_3-m_2)\Delta\cos(2{\rm Re}[\varrho])\right]\,.
\end{align}

\setcounter{equation}{0}
\section{Washout Strength in Resonant Leptogenesis with Two Sterile Neutrinos}
\label{appendix:washout:2RHN}

As for the equilibration of the sterile neutrinos, we note that
\begin{align}
{\rm tr}[YY^\dagger]\frac{\bar M}{8\pi H|_{T=\bar M}}\approx
108\cosh(2{\rm Im}[\varrho])
\,,
\end{align}
which can be inferred by substituting the observed neutrino masses (with $m_1=0$)
and mixing angles~\cite{Fogli:2012ua}
into Eq.~(\ref{epsbar:CI}). Using the relations~(\ref{eps:CI:onresonance})
or~(\ref{epsilons:multi:flavour}), it is clear that all eigenmodes are faster than
the Hubble expansion rate $H$ at the time when $T=\bar M$, what characterises strong washout.

For normal hierarchy, the $e$ flavour couples most weakly to the sterile neutrinos.
We find that
\begin{align}
[Y^\dagger Y]_{ee}=&
\frac{2\bar M}{v^2}
\Big(
\left[
m_3\sin^2\vartheta_{13}+m_2\sin^2\vartheta_{12}\cos^2\vartheta_{13}
\right]
\cosh(2{\rm Im}[\varrho])
\\\notag
-&2\sin\vartheta_{12}\sin\vartheta_{13}\cos\vartheta_{13}\sqrt{m_2 m_3}
\sin
\left(
\frac{\alpha_2}{2}+\delta
\right)
\sinh(2{\rm Im}[\varrho])
\Big)
\,,
\end{align}
where we parametrise the PMNS matrix as in Ref.~\cite{Drewes:2012ma}.
The washout strength has its global minimum along the curve where
$\sin(\alpha_2/2+\delta)=1$, where it is given by\footnote{The apparent
numerical coincidence in this equation is also noted in
Refs.~\cite{Shuve:2014zua,Asaka:2011pb}. If $\sin^2\vartheta_{13}$ ($\sin^2\vartheta_{12}$) were taking
values close to the upper (lower) observational bounds~\cite{Fogli:2012ua}, this may
lead to an even more effective suppression of the $e$ flavour from washout, with intersting consequences for
Leptogenesis from oscillations of sterile neutrinos~\cite{Drewes:2012ma,Akhmedov:1998qx,Asaka:2005pn,Shaposhnikov:2008pf,Canetti:2010aw,Canetti:2012vf,Canetti:2012zc,Canetti:2012kh,Shuve:2014zua}.}
\begin{align}
\frac{[Y^\dagger Y]_{ee}}{32\pi H|_{T=\bar M}}\approx
0.89\cosh(2{\rm Im}[\varrho])-0.84\sinh(2{\rm Im}[\varrho])
\end{align}
which takes for
${\rm Im}[\varrho]=0.87$ its minimum value $0.31$. Therefore, the $e$ flavour
will always equilibrate sufficiently long before freeze out at $z_{\rm f}={\cal O}(10)$.

\end{appendix}


\begin{thebibliography}{99}
\bibliographystyle{unsrt}

\bibitem{Beringer:1900zz}
  See, {\it e.g.} Chapter~12 of
  J.~Beringer {\it et al.}  [Particle Data Group Collaboration],
  ``Review of Particle Physics (RPP),''
  Phys.\ Rev.\ D {\bf 86} (2012) 010001.
  
\bibitem{Covi:1996wh}
  L.~Covi, E.~Roulet and F.~Vissani,
  ``CP violating decays in leptogenesis scenarios,''
  Phys.\ Lett.\ B {\bf 384} (1996) 169
  [hep-ph/9605319].
  
\bibitem{Flanz:1996fb}
  M.~Flanz, E.~A.~Paschos, U.~Sarkar and J.~Weiss,
  ``Baryogenesis through mixing of heavy Majorana neutrinos,''
  Phys.\ Lett.\ B {\bf 389} (1996) 693
  [hep-ph/9607310].

\bibitem{Pilaftsis:1997dr}
  A.~Pilaftsis,
  ``Resonant CP violation induced by particle mixing in transition amplitudes,''
  Nucl.\ Phys.\ B {\bf 504} (1997) 61
  [hep-ph/9702393].

\bibitem{Pilaftsis:1997jf}
  A.~Pilaftsis,
  ``CP violation and baryogenesis due to heavy Majorana neutrinos,''
  Phys.\ Rev.\ D {\bf 56} (1997) 5431
  [hep-ph/9707235].

\bibitem{Pilaftsis:2003gt}
  A.~Pilaftsis and T.~E.~J.~Underwood,
  ``Resonant leptogenesis,''
  Nucl.\ Phys.\ B {\bf 692} (2004) 303
  [hep-ph/0309342].

\bibitem{Pilaftsis:2005rv}
  A.~Pilaftsis and T.~E.~J.~Underwood,
  ``Electroweak-scale resonant leptogenesis,''
  Phys.\ Rev.\ D {\bf 72} (2005) 113001
  [hep-ph/0506107].


\bibitem{Giudice:2003jh}
  G.~F.~Giudice, A.~Notari, M.~Raidal, A.~Riotto and A.~Strumia,
  ``Towards a complete theory of thermal leptogenesis in the SM and MSSM,''
  Nucl.\ Phys.\ B {\bf 685} (2004) 89
  [hep-ph/0310123].
  
\bibitem{Buchmuller:2004nz}
  W.~Buchm\"uller, P.~Di Bari and M.~Pl\"umacher,
  ``Leptogenesis for pedestrians,''
  Annals Phys.\  {\bf 315} (2005) 305
  [hep-ph/0401240].
  
\bibitem{Kolb:1979qa}
  E.~W.~Kolb and S.~Wolfram,
  ``Baryon Number Generation in the Early Universe,''
  Nucl.\ Phys.\ B {\bf 172} (1980) 224
   [Erratum-ibid.\ B {\bf 195} (1982) 542].

\bibitem{Schwinger:1960qe}
  J.~S.~Schwinger,
  ``Brownian motion of a quantum oscillator,''
  J.\ Math.\ Phys.\  {\bf 2} (1961) 407.

\bibitem{Keldysh:1964ud}
  L.~V.~Keldysh,
  ``Diagram technique for nonequilibrium processes,''
  Zh.\ Eksp.\ Teor.\ Fiz.\  {\bf 47} (1964) 1515
  [Sov.\ Phys.\ JETP {\bf 20} (1965) 1018].

\bibitem{Calzetta:1986cq}
  E.~Calzetta and B.~L.~Hu,
  ``Nonequilibrium Quantum Fields: Closed Time Path Effective Action, Wigner
  Function and Boltzmann Equation,''
  Phys.\ Rev.\  D {\bf 37} (1988) 2878.

\bibitem{Buchmuller:2000nd} 
  W.~Buchmuller and S.~Fredenhagen,
  ``Quantum mechanics of baryogenesis,''
  Phys.\ Lett.\ B {\bf 483}, 217 (2000)
  [hep-ph/0004145].

\bibitem{De Simone:2007rw}
  A.~De Simone and A.~Riotto,
  ``Quantum Boltzmann Equations and Leptogenesis,''
  JCAP {\bf 0708} (2007) 002
  [hep-ph/0703175].

\bibitem{Garny:2009rv}
  M.~Garny, A.~Hohenegger, A.~Kartavtsev and M.~Lindner,
  ``Systematic approach to leptogenesis in nonequilibrium QFT: vertex
  contribution to the CP-violating parameter,''
  Phys.\ Rev.\  D {\bf 80} (2009) 125027
  [arXiv:0909.1559 [hep-ph]].

\bibitem{Garny:2009qn}
  M.~Garny, A.~Hohenegger, A.~Kartavtsev and M.~Lindner,
  ``Systematic approach to leptogenesis in nonequilibrium QFT: self-energy
  contribution to the CP-violating parameter,''
  Phys.\ Rev.\  D {\bf 81} (2010) 085027
  [arXiv:0911.4122 [hep-ph]].

\bibitem{Anisimov:2010aq}
  A.~Anisimov, W.~Buchm\"uller, M.~Drewes and S.~Mendizabal,
  ``Leptogenesis from Quantum Interference in a Thermal Bath,''
  Phys.\ Rev.\ Lett.\  {\bf 104} (2010) 121102
  [arXiv:1001.3856 [hep-ph]].

\bibitem{Garny:2010nj}
  M.~Garny, A.~Hohenegger, A.~Kartavtsev,
  ``Medium corrections to the CP-violating parameter in leptogenesis,''
  Phys.\ Rev.\  {\bf D81 } (2010)  085028.
  [arXiv:1002.0331 [hep-ph]].

\bibitem{Beneke:2010wd}
  M.~Beneke, B.~Garbrecht, M.~Herranen and P.~Schwaller,
  ``Finite Number Density Corrections to Leptogenesis,''
  Nucl.\ Phys.\  B {\bf 838} (2010) 1
  [arXiv:1002.1326 [hep-ph]].

\bibitem{Beneke:2010dz}
  M.~Beneke, B.~Garbrecht, C.~Fidler, M.~Herranen and P.~Schwaller,
  ``Flavoured Leptogenesis in the CTP Formalism,''
  Nucl.\ Phys.\  B {\bf 843} (2011) 177
  [arXiv:1007.4783 [hep-ph]].

\bibitem{Garny:2010nz}
  M.~Garny, A.~Hohenegger and A.~Kartavtsev,
  ``Quantum corrections to leptogenesis from the gradient expansion,''
  arXiv:1005.5385 [hep-ph].

\bibitem{Anisimov:2010dk}
  A.~Anisimov, W.~Buchmuller, M.~Drewes and S.~Mendizabal,
  ``Quantum Leptogenesis I,''
  Annals Phys.\  {\bf 326} (2011) 1998
  [arXiv:1012.5821 [hep-ph]].


\bibitem{Garbrecht:2011aw}
  B.~Garbrecht and M.~Herranen,
  ``Effective Theory of Resonant Leptogenesis in the Closed-Time-Path Approach,''
  Nucl.\ Phys.\ B {\bf 861} (2012) 17
  [arXiv:1112.5954 [hep-ph]].

\bibitem{Drewes:2012ma}
  M.~Drewes and B.~Garbrecht,
  ``Leptogenesis from a GeV Seesaw without Mass Degeneracy,''
  JHEP {\bf 1303} (2013) 096
  [arXiv:1206.5537 [hep-ph]].
  
\bibitem{Garny:2011hg}
  M.~Garny, A.~Kartavtsev and A.~Hohenegger,
  ``Leptogenesis from first principles in the resonant regime,''
  Annals Phys.\  {\bf 328} (2013) 26
  [arXiv:1112.6428 [hep-ph]].

\bibitem{Iso:2013lba}
  S.~Iso, K.~Shimada and M.~Yamanaka,
  ``Kadanoff-Baym approach to the thermal resonant leptogenesis,''
  JHEP {\bf 1404} (2014) 062
  [arXiv:1312.7680 [hep-ph]].
  
\bibitem{Iso:2014afa}
  S.~Iso and K.~Shimada,
  ``Coherent Flavour Oscillation and CP Violating Parameter in Thermal Resonant Leptogenesis,''
  arXiv:1404.4816 [hep-ph].

\bibitem{Hohenegger:2014cpa}
  A.~Hohenegger and A.~Kartavtsev,
  ``Leptogenesis in crossing and runaway regimes,''
  arXiv:1404.5309 [hep-ph].
  
\bibitem{Akhmedov:1998qx}
  E.~K.~Akhmedov, V.~A.~Rubakov and A.~Y.~.Smirnov,
  ``Baryogenesis via neutrino oscillations,''
  Phys.\ Rev.\ Lett.\  {\bf 81} (1998) 1359
  [hep-ph/9803255].
  
\bibitem{Asaka:2005pn}
  T.~Asaka and M.~Shaposhnikov,
  ``The nuMSM, dark matter and baryon asymmetry of the universe,''
  Phys.\ Lett.\ B {\bf 620} (2005) 17
  [hep-ph/0505013].

\bibitem{Shaposhnikov:2008pf}
  M.~Shaposhnikov,
  ``The nuMSM, leptonic asymmetries, and properties of singlet fermions,''
  JHEP {\bf 0808} (2008) 008
  [arXiv:0804.4542 [hep-ph]].


\bibitem{Canetti:2010aw}
  L.~Canetti and M.~Shaposhnikov,
  ``Baryon Asymmetry of the Universe in the NuMSM,''
  JCAP {\bf 1009} (2010) 001
  [arXiv:1006.0133 [hep-ph]].
 
 
\bibitem{Canetti:2012vf}
  L.~Canetti, M.~Drewes and M.~Shaposhnikov,
  ``Sterile Neutrinos as the Origin of Dark and Baryonic Matter,''
  Phys.\ Rev.\ Lett.\  {\bf 110} (2013) 6,  061801
  [arXiv:1204.3902 [hep-ph]].
  
\bibitem{Canetti:2012zc}
  L.~Canetti, M.~Drewes and M.~Shaposhnikov,
  ``Matter and Antimatter in the Universe,''
  New J.\ Phys.\  {\bf 14} (2012) 095012
  [arXiv:1204.4186 [hep-ph]].



\bibitem{Canetti:2012kh}
  L.~Canetti, M.~Drewes, T.~Frossard and M.~Shaposhnikov,
  ``Dark Matter, Baryogenesis and Neutrino Oscillations from Right Handed Neutrinos,''
  Phys.\ Rev.\ D {\bf 87} (2013) 9,  093006
  [arXiv:1208.4607 [hep-ph]].
  

\bibitem{Shuve:2014zua}
  B.~Shuve and I.~Yavin,
  ``Baryogenesis through Neutrino Oscillations: A Unified Perspective,''
  Phys.\ Rev.\ D {\bf 89} (2014) 075014
  [arXiv:1401.2459 [hep-ph]].
  
  
\bibitem{Dev:2014laa}
  P.~S.~B.~Dev, P.~Millington, A.~Pilaftsis and D.~Teresi,
  ``Flavour Covariant Transport Equations: an Application to Resonant Leptogenesis,''
  arXiv:1404.1003 [hep-ph].
  
\bibitem{Cirigliano:2009yt}
  V.~Cirigliano, C.~Lee, M.~J.~Ramsey-Musolf and S.~Tulin,
  ``Flavored Quantum Boltzmann Equations,''
  Phys.\ Rev.\ D {\bf 81} (2010) 103503
  [arXiv:0912.3523 [hep-ph]].
  
  
\bibitem{Sigl:1992fn}
  G.~Sigl and G.~Raffelt,
  ``General kinetic description of relativistic mixed neutrinos,''
  Nucl.\ Phys.\ B {\bf 406} (1993) 423.
  
\bibitem{Fidler:2011yq}
  C.~Fidler, M.~Herranen, K.~Kainulainen and P.~M.~Rahkila,
  ``Flavoured quantum Boltzmann equations from cQPA,''
  JHEP {\bf 1202} (2012) 065
  [arXiv:1108.2309 [hep-ph]].


\bibitem{Blanchet:2011xq}
  S.~Blanchet, P.~Di Bari, D.~A.~Jones and L.~Marzola,
  ``Leptogenesis with heavy neutrino flavours: from density matrix to Boltzmann equations,''
  JCAP {\bf 1301} (2013) 041
  [arXiv:1112.4528 [hep-ph]].


  
\bibitem{Blanchet:2006be}
  S.~Blanchet and P.~Di Bari,
  ``Flavor effects on leptogenesis predictions,''
  JCAP {\bf 0703} (2007) 018
  [hep-ph/0607330].

  
\bibitem{Garbrecht:2012pq}
  B.~Garbrecht,
  ``Baryogenesis from Mixing of Lepton Doublets,''
  Nucl.\ Phys.\ B {\bf 868} (2013) 557
  [arXiv:1210.0553 [hep-ph]].

\bibitem{Kolb:1983ni}
  E.~W.~Kolb and M.~S.~Turner,
  ``Grand Unified Theories and the Origin of the Baryon Asymmetry,''
  Ann.\ Rev.\ Nucl.\ Part.\ Sci.\  {\bf 33} (1983) 645.
  
\bibitem{Abada:2006fw}
  A.~Abada, S.~Davidson, F.~X.~Josse-Michaux, M.~Losada and A.~Riotto,
  ``Flavour Issues in Leptogenesis,''
  JCAP {\bf 0604}, 004 (2006)
  [arXiv:hep-ph/0601083].

\bibitem{Nardi:2006fx}
  E.~Nardi, Y.~Nir, E.~Roulet and J.~Racker,
  ``The importance of flavor in leptogenesis,''
  JHEP {\bf 0601}, 164 (2006)
  [arXiv:hep-ph/0601084].
  
\bibitem{Antusch:2010ms}
  S.~Antusch, P.~Di Bari, D.~A.~Jones and S.~F.~King,
  ``A fuller flavour treatment of $N_2$-dominated leptogenesis,''
  Nucl.\ Phys.\ B {\bf 856} (2012) 180
  [arXiv:1003.5132 [hep-ph]].


\bibitem{Casas:2001sr}
  J.~A.~Casas and A.~Ibarra,
  ``Oscillating neutrinos and $\mu \to e \gamma$,''
  Nucl.\ Phys.\ B {\bf 618} (2001) 171
  [hep-ph/0103065].

  
\bibitem{Fogli:2012ua}
  G.~L.~Fogli, E.~Lisi, A.~Marrone, D.~Montanino, A.~Palazzo and A.~M.~Rotunno,
  ``Global analysis of neutrino masses, mixings and phases: entering the era of leptonic CP violation searches,''
  Phys.\ Rev.\ D {\bf 86} (2012) 013012
  [arXiv:1205.5254 [hep-ph]].
  
\bibitem{Forero:2014bxa}
  D.~V.~Forero, M.~Tortola and J.~W.~F.~Valle,
  ``Neutrino oscillations refitted,''
  arXiv:1405.7540 [hep-ph].

  
\bibitem{Barbieri:1999ma}
  R.~Barbieri, P.~Creminelli, A.~Strumia and N.~Tetradis,
  ``Baryogenesis through leptogenesis,''
  Nucl.\ Phys.\ B {\bf 575} (2000) 61
  [hep-ph/9911315].
  
\bibitem{Buchmuller:2001sr}
  W.~Buchm\"uller and M.~Pl\"umacher,
  ``Spectator processes and baryogenesis,''
  Phys.\ Lett.\ B {\bf 511} (2001) 74
  [hep-ph/0104189].

  
\bibitem{Davidson:2008bu}
  S.~Davidson, E.~Nardi and Y.~Nir,
  ``Leptogenesis,''
  Phys.\ Rept.\  {\bf 466} (2008) 105
  [arXiv:0802.2962 [hep-ph]].
  
\bibitem{Salvio:2011sf}
  A.~Salvio, P.~Lodone and A.~Strumia,
  ``Towards leptogenesis at NLO: the right-handed neutrino interaction rate,''
  JHEP {\bf 1108} (2011) 116
  [arXiv:1106.2814 [hep-ph]].
  
\bibitem{Laine:2011pq}
  M.~Laine and Y.~Schroder,
  ``Thermal right-handed neutrino production rate in the non-relativistic regime,''
  JHEP {\bf 1202} (2012) 068
  [arXiv:1112.1205 [hep-ph]].
  
\bibitem{Biondini:2013xua}
  S.~Biondini, N.~Brambilla, M.~A.~Escobedo and A.~Vairo,
  ``An effective field theory for non-relativistic Majorana neutrinos,''
  JHEP {\bf 1312} (2013) 028
  [arXiv:1307.7680, arXiv:1307.7680].

  
\bibitem{Garbrecht:2013gd}
  B.~Garbrecht, F.~Glowna and M.~Herranen,
  ``Right-Handed Neutrino Production at Finite Temperature: Radiative Corrections, Soft and Collinear Divergences,''
  JHEP {\bf 1304} (2013) 099
  [arXiv:1302.0743 [hep-ph]].
  
\bibitem{Laine:2013lka}
  M.~Laine,
  ``Thermal right-handed neutrino production rate in the relativistic regime,''
  JHEP {\bf 1308} (2013) 138
  [arXiv:1307.4909 [hep-ph]].

\bibitem{Bodeker:2014hqa}
  D.~Bodeker and M.~Laine,
  ``Kubo relations and radiative corrections for lepton number washout,''
  JCAP {\bf 1405} (2014) 041
  [arXiv:1403.2755 [hep-ph]].
  
\bibitem{Anisimov:2005hr}
  A.~Anisimov, A.~Broncano and M.~Plumacher,
  ``The CP-asymmetry in resonant leptogenesis,''
  Nucl.\ Phys.\ B {\bf 737} (2006) 176
  [hep-ph/0511248].

\bibitem{Asaka:2011pb}
  T.~Asaka, S.~Eijima and H.~Ishida,
  ``Mixing of Active and Sterile Neutrinos,''
  JHEP {\bf 1104} (2011) 011
  [arXiv:1101.1382 [hep-ph]].
  
\bibitem{Garbrecht:2014kda}
  B.~Garbrecht and P.~Schwaller,
  ``Spectator Effects during Leptogenesis in the Strong Washout Regime,''
  arXiv:1404.2915 [hep-ph].


\end{thebibliography}
\end{document}